\batchmode
\makeatletter
\def\input@path{{"/home/jacob/Documents/Work/My Papers/Manifestly Covariant Lagrangians (2020)/"}}
\makeatother
\documentclass[twoside,twocolumn,english,superscriptaddress,nobibnotes,nofootinbib]{revtex4-2}
\usepackage[latin9]{inputenc}
\usepackage{color}
\usepackage{babel}
\usepackage{mathrsfs}
\usepackage{mathtools}
\usepackage{amsmath}
\usepackage{amssymb}
\usepackage[pdftex,unicode=true,
 bookmarks=true,bookmarksnumbered=true,bookmarksopen=false,
 breaklinks=false,pdfborder={0 0 0},pdfborderstyle={},backref=false,colorlinks=true]
 {hyperref}
\hypersetup{pdftitle={Manifestly Covariant Lagrangians, Classical Particles with Spin, and the Origins of Gauge Invariance},
 pdfauthor={Jacob A. Barandes},
 pdfsubject={Physics},
 pdfkeywords={Classical physics; particle physics; gauge theories; analytical dynamics},
 linkcolor=blue, urlcolor=blue, citecolor=blue}

\makeatletter

\usepackage{url}

\let\originalleft\left
\let\originalright\right
\renewcommand{\left}{\mathopen{}\mathclose\bgroup\originalleft}
\renewcommand{\right}{\aftergroup\egroup\originalright}

\def\smalloverbrace#1{\mathop{\vbox{\m@th\ialign{##\crcr%
      \noalign{\kern3\p@}%
      \tiny\downbracefill\crcr\noalign{\kern3\p@\nointerlineskip}%
      $\hfil\displaystyle{#1}\hfil$\crcr}}}\limits}

\def\smallunderbrace#1{\mathop{\vtop{\m@th\ialign{##\crcr
   $\hfil\displaystyle{#1}\hfil$\crcr
   \noalign{\kern3\p@\nointerlineskip}%
   \tiny\upbracefill\crcr\noalign{\kern3\p@}}}}\limits}

\makeatother

\begin{document}
\title{Manifestly Covariant Lagrangians, Classical Particles with Spin, and
the Origins of Gauge Invariance}
\author{Jacob A. Barandes}
\email{jacob\_barandes@harvard.edu}

\affiliation{Jefferson Physical Laboratory, Harvard University, Cambridge, MA 02138}
\date{\today}
\begin{abstract}
In this paper, we review a general technique for converting the standard
Lagrangian description of a classical system into a formulation that
puts time on an equal footing with the system's degrees of freedom.
We show how the resulting framework anticipates key features of special
relativity, including the signature of the Minkowski metric tensor
and the special role played by theories that are invariant under a
generalized notion of Lorentz transformations. We then use this technique
to revisit a classification of classical particle-types that mirrors
Wigner's classification of quantum particle-types in terms of irreducible
representations of the Poincaré group, including the cases of massive
particles, massless particles, and tachyons. Along the way, we see
gauge invariance naturally emerge in the context of classical massless
particles with nonzero spin, as well as study the massless limit of
a massive particle and derive a classical-particle version of the
Higgs mechanism.
\end{abstract}
\maketitle

\global\long\def\vec#1{{\bf #1}}%
\global\long\def\vecgreek#1{\boldsymbol{#1}}%
\global\long\def\dotprod{\cdot}%
\global\long\def\crossprod{\times}%
\global\long\def\tud#1#2#3{#1^{#2}{}_{#3}}%
\global\long\def\tdu#1#2#3{#1_{#2}{}^{#3}}%
\global\long\def\defeq{\equiv}%
\global\long\def\Trace{\mathrm{Tr}}%
\global\long\def\transp{\mathrm{T}}%
\global\long\def\refvalue{0}%
\global\long\def\parens#1{(#1)}%
\global\long\def\bigparens#1{\big(#1\big)}%
\global\long\def\Bigparens#1{\Big(#1\Big)}%
\global\long\def\biggparens#1{\bigg(#1\bigg)}%
\global\long\def\Biggparens#1{\Bigg(#1\Bigg)}%
\global\long\def\bracks#1{[#1]}%
\global\long\def\bigbracks#1{\big[#1\big]}%
\global\long\def\Bigbracks#1{\Big[#1\Big]}%
\global\long\def\biggbracks#1{\bigg[#1\bigg]}%
\global\long\def\Biggbracks#1{\Bigg[#1\Bigg]}%
\global\long\def\curlies#1{\{#1\}}%
\global\long\def\bigcurlies#1{\big\{#1\big\}}%
\global\long\def\Bigcurlies#1{\Big\{#1\Big\}}%
\global\long\def\biggcurlies#1{\bigg\{#1\bigg\}}%
\global\long\def\Biggcurlies#1{\Bigg\{#1\Bigg\}}%
\global\long\def\verts#1{\vert#1\vert}%
\global\long\def\bigverts#1{\big\vert#1\big\vert}%
\global\long\def\Bigverts#1{\Big\vert#1\Big\vert}%
\global\long\def\biggverts#1{\bigg\vert#1\bigg\vert}%
\global\long\def\Biggverts#1{\Bigg\vert#1\Bigg\vert}%

\section{Introduction}

The Lagrangian formulation of classical physics provides an elegant
and powerful set of techniques for analyzing the behavior of physical
systems. For classical fields, it is customary to employ Lagrangians
that make the symmetries of special relativity manifest, but textbook
treatments of mechanical systems tend to treat time and energy very
differently from degrees of freedom and momenta.

In this paper,\footnote{For a synopsis of the results obtained in this paper, see \citep{Barandes:2021gifcmpws}.}
we cast new light on a technique for resolving this shortcoming. Among
its useful features, we show that this framework anticipates key aspects
of special relativity, like the signature of the Minkowski metric
tensor and the special role played by classical systems that exhibit
generalizations of Lorentz invariance.

Extending earlier work, including \citep{SudarshanMukunda:1974cdamp,BalachandranMarmoSkagerstamStern:1983gsfb,Souriau:1997sds,Rivas:2002ktsp},
we then present a fully classical version of Wigner's famous classification
\citep{Wigner:1939urilg} of quantum particles into general types\textemdash massive,
massless, and tachyonic. In close parallel with Wigner's construction,
which is based on identifying the Hilbert spaces of quantum particles
with irreducible representations of the Poincaré group, our classification
of classical particle-types consists of identifying their phase spaces
with ``irreducible'' (or, more properly, transitive) group actions
of the Poincaré group, so that those phase spaces serve as homogeneous
spaces for the Poincaré group. Our classical particles generically
possess fixed total spin, but without spin quantization, and therefore
correspond to the limit of large spin quantum numbers.

Along the way, and as a case study in how kinematics can determine
dynamics, we show that the structure of these phase spaces leads to
a simple Lagrangian formulation that can handle both massive and massless
particles, and that neatly accommodates spin. In addition, by paying
careful attention to the compactness properties of these phase spaces
at fixed energy, we show that physically acceptable massless particles
with spin feature a classical point-particle manifestation of gauge
invariance that is deeply connected to the gauge invariance of electromagnetism\textemdash meaning
that this form of gauge invariance is not solely a property of classical
field theory or of relativistic quantum mechanics. By studying the
relationship between the massive and massless cases through the massless
limit, we also derive a classical point-particle version of the Higgs
mechanism.

\section{The Lagrangian Formulation}

We start with a brief review of general classical systems and their
standard Lagrangian formulation.\footnote{For a more extensive introduction, see \citep{GoldsteinSafkoPoole:2013cm}.}
Afterward, we will turn to the development of a manifestly covariant
approach.

\subsection{Classical Systems}

In general, the mathematical description of a classical system consists
of a configuration space whose points denote the possible ``snapshots''
that the system can occupy, together with a list of rules or laws
that determine how the system's instantaneous configuration is allowed
to evolve.

If $q_{\alpha}$ are a collection of independent numerical coordinates
that label the points in the system's configuration space, with $\alpha$
an index distinguishing the different coordinates, then we call $q_{\alpha}$
a set of degrees of freedom for the system. We will assume for simplicity
that we can cover the entire configuration space with a single such
coordinate system, apart from possible regions of measure zero where
the coordinates are not well-defined.

A candidate trajectory of the system is an arbitrary smooth path
through the system's configuration space, and is conveniently defined
by specifying the system's degrees of freedom $q_{\alpha}\parens t$
as functions of a real-valued parameter $t$ called the time. The
system's rates of change are then denoted by $\dot{q}_{\alpha}\parens t$,
where dots denote derivatives with respect to $t$: 
\begin{align}
\dot{q}_{\alpha}\parens t & \defeq\frac{dq_{\alpha}\parens t}{dt},\label{eq:GeneralDefRatesChangeDOF}\\
\ddot{q}_{\alpha}\parens t & \defeq\frac{d^{2}q_{\alpha}\parens t}{dt^{2}},\label{eq:GeneralDefAccelsDOF}
\end{align}
 and so forth. Altogether, the system's configuration space, a choice
of degrees of freedom $q_{\alpha}$, and all the system's candidate
trajectories make up the system's kinematics.

The rules or laws that prescribe which \emph{candidate} trajectories
are \emph{physical} trajectories that the system can actually follow
make up the system's dynamics. In the simplest cases, these rules
take the form of first- or second-order differential equations of
the form 
\begin{equation}
f_{\alpha}\parens{q,\dot{q},\ddot{q}}=0,\label{eq:GeneralEOM}
\end{equation}
 which are called the system's equations of motion.

As a simple example, consider a Newtonian particle of constant inertial
mass $m$ in an inertial reference frame in three spatial dimensions.
At the level of kinematics, the particle has a three-dimensional configuration
space isomorphic to $\mathbb{R}^{3}$, and three degrees of freedom
$q_{x},q_{y},q_{z}$ that make up the particle's position vector $\vec X$
in Cartesian coordinates: 
\begin{equation}
\vec X\defeq\parens{X,Y,Z}\defeq\parens{q_{x},q_{y},q_{z}}.\label{eq:3DNewtonianParticleDOF}
\end{equation}
 At the level of dynamics, we assume a given force vector 
\begin{equation}
\vec F\defeq\parens{F_{x},F_{y},F_{z}},\label{eq:3DNewtonianParticleForceVector}
\end{equation}
 in which case the system's equations of motion make up the three
components of Newton's second law, 
\begin{equation}
\vec F=m\vec a,\label{eq:3DNewtonsSecondLaw}
\end{equation}
 where $\vec a$ is the system's acceleration vector: 
\begin{equation}
\vec a\defeq\ddot{\vec X}=\parens{\ddot{X},\ddot{Y},\ddot{Z}}.\label{eq:3DNewtonianAccelerationVector}
\end{equation}

\subsection{The Lagrangian Formulation}

Returning again to the case of a general classical system, let $L\parens{q,\dot{q},t}$,
assumed to have units of energy, be a function of the system's degrees
of freedom $q_{\alpha}$, its rates of change $\dot{q}_{\alpha}$,
and the time $t$. Notice that if we do not specify a candidate trajectory,
then $q_{\alpha}$, $\dot{q}_{\alpha}$, and $t$ are all independent
variables.

By contrast, if we are given a candidate trajectory $q_{\alpha}\parens t$
from an arbitrary initial time $t_{A}$ to an arbitrary final time
$t_{B}$, then the degrees of freedom $q_{\alpha}\parens t$ and their
rates of change $\dot{q}_{\alpha}\parens t$ become functions of $t$,
and we can define an integral of $L\parens{q\parens t,\dot{q}\parens t,t}$
over time: 
\begin{equation}
S\bracks q\defeq\int_{t_{A}}^{t_{B}}\negthickspace\negthickspace dt\,L\parens{q\parens t,\dot{q}\parens t,t}.\label{eq:ClassicalActionFromLagrangian}
\end{equation}
 The bracketed argument $\bracks q$ in this notation indicates that
$S\bracks q$ is a functional of the system's candidate trajectory,
meaning that $S\bracks q$ depends on the infinite continuum of real
numbers\textemdash parametrized by $t$\textemdash that make up the
entire candidate trajectory $q_{\alpha}\parens t$.

If we extremize $S\bracks q$ over all candidate trajectories that
share the same initial and final conditions, 
\begin{align}
 & \delta S\bracks q=0,\nonumber \\
 & \quad\textrm{with \ensuremath{q_{\alpha}\parens{t_{A}}} and \ensuremath{q_{\alpha}\parens{t_{B}}} held fixed for all \ensuremath{\alpha},}\label{eq:VariationActionFunctional}
\end{align}
 then, as we will review in detail, we obtain the Euler-Lagrange equations,
\begin{equation}
\frac{\partial L}{\partial q_{\alpha}}-\frac{d}{dt}\biggparens{\frac{\partial L}{\partial\dot{q}_{\alpha}}}=0.\label{eq:EulerLagrangeEquations}
\end{equation}
 For many Lagrangians used in practice, the Euler-Lagrange equations
are typically second-order in the time $t$. If the Euler-Lagrange
equations collectively turn out to be equivalent to the system's equations
of motion \eqref{eq:GeneralEOM}, then we respectively call $L=L\parens{q,\dot{q},t}$
and $S\bracks q$ a Lagrangian and an action functional for the
system, and we say that $S\bracks q\defeq\int dt\,L$ provides a Lagrangian formulation
for the system. (Note that $L$ and $S\bracks q$ are generally not
unique for a given system.)

Deriving the Euler-Lagrange equations from the extremization condition
\eqref{eq:VariationActionFunctional}, known as Hamilton's principle
or the principle of least action, takes just a few steps, and will
be an illustrative exercise before we generalize the construction
later on. We start by varying the system's candidate trajectory $q_{\alpha}\parens t$
according to 
\begin{equation}
q_{\alpha}\parens t\mapsto q_{\alpha}\parens t+\delta q_{\alpha}\parens t,\label{eq:ActionPrincipleTrajectoryVariation}
\end{equation}
 where the variations $\delta q_{\alpha}\parens t$ are infinitesimal
functions of the time $t$ that are assumed to vanish at the endpoints
of the system's trajectory in keeping with \eqref{eq:VariationActionFunctional},
\begin{equation}
\delta q_{\alpha}\parens{t_{A}}=0,\quad\delta q_{\alpha}\parens{t_{B}}=0,\label{eq:ActionPrincipleEndpointsFixed}
\end{equation}
 but are otherwise arbitrary and independent. Taking a time derivative
of the variation rule \eqref{eq:ActionPrincipleTrajectoryVariation}
yields the corresponding variations in the system's rates of change
$\dot{q}_{\alpha}\parens t$: 
\begin{align}
\dot{q}_{\alpha}\parens t=\frac{dq_{\alpha}\parens t}{dt} & \mapsto\frac{d\parens{q_{\alpha}\parens t+\delta q_{\alpha}\parens t}}{dt}\nonumber \\
 & \quad=\dot{q}_{\alpha}\parens t+\frac{d}{dt}\delta q_{\alpha}\parens t.\label{eq:ActionPrincipleTrajectoryVariationRates}
\end{align}
 We infer that the induced variation in $\dot{q}_{\alpha}\parens t$
is precisely the time derivative of the variation in $q_{\alpha}\parens t$,
\begin{equation}
\delta\dot{q}_{\alpha}\parens t=\frac{d}{dt}\delta q_{\alpha}\parens t,\label{eq:ActionPrincipleTimeDerivVariationCommute}
\end{equation}
 so, loosely speaking, the variation operator $\delta$ ``commutes''
with the time derivative $d/dt$.

Applying the extremization condition \eqref{eq:VariationActionFunctional},
using the chain rule, carrying out an integration by parts, and dropping
boundary terms that vanish by the assumption that the variations vanish
at the initial and final times, we find 
\begin{align}
\delta S\bracks q & \defeq\int dt\,L\parens{q+\delta q,\dot{q}+\delta\dot{q},t}\nonumber \\
 & \qquad\qquad-\int dt\,L\parens{q,\dot{q},t}\nonumber \\
 & =\int dt\,\sum_{\alpha}\biggparens{\frac{\partial L}{\partial q_{\alpha}}\delta q_{\alpha}+\frac{\partial L}{\partial\dot{q}_{\alpha}}\delta\dot{q}_{\alpha}}\nonumber \\
 & =\int dt\,\sum_{\alpha}\biggparens{\frac{\partial L}{\partial q_{\alpha}}\delta q_{\alpha}+\frac{\partial L}{\partial\dot{q}_{\alpha}}\frac{d}{dt}\delta q_{\alpha}}\nonumber \\
 & =\int dt\,\sum_{\alpha}\biggparens{\frac{\partial L}{\partial q_{\alpha}}-\frac{d}{dt}\biggparens{\frac{\partial L}{\partial\dot{q}_{\alpha}}}}\delta q_{\alpha}=0.\label{eq:VariationActionDerivationEulerLagrange}
\end{align}
 Because the infinitesimal variations $\delta q_{\alpha}\parens t$
are assumed to be arbitrary and independent within the domain of integration,
we conclude that the factor in parentheses must be zero, so we end
up with the Euler-Lagrange equations \eqref{eq:EulerLagrangeEquations},
as claimed.

As an example, consider a Newtonian particle of constant inertial
mass $m$ and position vector $\vec X\defeq\parens{X,Y,Z}$ with kinetic
energy 
\begin{equation}
T\parens{\dot{\vec X}}=\frac{1}{2}m\dot{\vec X}^{2}=\frac{1}{2}m\parens{\dot{X}^{2}+\dot{Y}^{2}+\dot{Z}^{2}}\label{eq:3DNewtonianParticleKineticEnergy}
\end{equation}
 and subject to a conservative force 
\begin{equation}
\vec F=-\nabla V=\biggparens{-\frac{\partial V}{\partial X},-\frac{\partial V}{\partial Y},-\frac{\partial V}{\partial Z}},\label{eq:NewtonianForceFromGradPotential}
\end{equation}
 corresponding to a potential energy $V\parens{\vec X}=V\parens{X,Y,Z}$.
If we choose the Lagrangian 
\begin{equation}
L\parens{\vec X,\dot{\vec X}}\defeq T-V=\frac{1}{2}m\dot{\vec X}^{2}-V\parens{\vec X},\label{eq:NewtonianLagrangian}
\end{equation}
 then the Euler-Lagrange equations \eqref{eq:EulerLagrangeEquations},
with $\vec X=\parens{X,Y,Z}=\parens{q_{x},q_{y},q_{z}}$, give 
\begin{align*}
 & \frac{\partial L}{\partial X_{i}}-\frac{d}{dt}\biggparens{\frac{\partial L}{\partial\dot{X}_{i}}}\\
 & \qquad=-\frac{\partial V}{\partial X_{i}}-m\ddot{X}_{i}=0,
\end{align*}
 which replicate the three components of Newton's second law \eqref{eq:3DNewtonsSecondLaw},
$\vec F=m\vec a$. Notice also that the object's momentum 
\begin{equation}
\vec p\defeq\parens{p_{x},p_{y},p_{z}}\defeq m\dot{\vec X}\label{eq:3DNewtonianMomentum}
\end{equation}
 is related to the Lagrangian \eqref{eq:NewtonianLagrangian} by 
\begin{equation}
p_{i}=m\dot{X}_{i}=\frac{\partial L}{\partial\dot{X}_{i}},\label{eq:NewtonianCanonicalMomenta}
\end{equation}
 and that the object's total mechanical energy 
\begin{equation}
E\defeq T+V\label{eq:NewtonianMechanicalEnergy}
\end{equation}
 is related to $\vec p$ and $L$ by 
\begin{align}
E & =\frac{1}{2}m\dot{\vec X}^{2}+V\parens{\vec X}=\frac{\vec p^{2}}{2m}+V\parens{\vec X}\nonumber \\
 & =\vec p\dotprod\dot{\vec X}-L.\label{eq:DefNewtonianEnergyAsHamiltonian}
\end{align}

For a generic physical system that may not resemble a Newtonian object,
we might not have an obvious choice for defining the system's momenta
and energy. The formulas at the end of \eqref{eq:NewtonianCanonicalMomenta}
and at the end of \eqref{eq:DefNewtonianEnergyAsHamiltonian} have
the virtue of being general and of leading to quantities $p_{i}$
and $E$ that, as we will see shortly, are respectively conserved
if the system's action functional \eqref{eq:ClassicalActionFromLagrangian}
is symmetric under translations in space, $X_{i}\mapsto X_{i}+\parens{\textrm{constant}}$,
or under translations in time, $t\mapsto t+\parens{\textrm{constant}}$.

Given a generic system with a Lagrangian formulation, we are therefore
motivated to define each of the system's canonical momenta $p_{\alpha}$
in terms of the system's Lagrangian $L$ as the partial derivative
of $L$ with respect to the corresponding rate of change $\dot{q}_{\alpha}$:
\begin{equation}
p_{\alpha}\defeq\frac{\partial L}{\partial\dot{q}_{\alpha}}.\label{eq:DefCanonicalMomenta}
\end{equation}
Recalling that the set of points labeled by particular values $q_{\alpha}$
of a system's degrees of freedom define the system's configuration
space, the set of points $\parens{q,p}$ labeled by particular values
of the system's canonical variables $q_{\alpha}$ and $p_{\alpha}$
define the system's phase space.

If we can solve the definitions \eqref{eq:DefCanonicalMomenta} for
the rates of change $\dot{q}_{\alpha}$ as functions of the canonical
variables $q_{\alpha}$ and $p_{\alpha}$, then the system's Hamiltonian
$H\parens{q,p,t}$, which is a function on the system's phase space
and roughly describes the system's energy, is defined as 
\begin{align}
H & \defeq\sum_{\alpha}\frac{\partial L}{\partial\dot{q}_{\alpha}}\dot{q}_{\alpha}-L,\nonumber \\
 & =\sum_{\alpha}p_{\alpha}\dot{q}_{\alpha}-L,\label{eq:DefHamiltonian}
\end{align}
 which is known as a Legendre transformation of $L$.

In terms of the canonical momenta \eqref{eq:DefCanonicalMomenta},
we can recast the Euler-Lagrange equations \eqref{eq:EulerLagrangeEquations}
as 
\begin{equation}
\frac{dp_{\alpha}}{dt}=\frac{\partial L}{\partial q_{\alpha}}.\label{eq:TimeDerivCanonicalMomFromLagr}
\end{equation}
 One can also use the chain rule together with the Euler-Lagrange
equations to show that 
\begin{align*}
\frac{dH}{dt} & =\sum_{\alpha}\dot{p}_{\alpha}\dot{q}_{\alpha}+\sum_{\alpha}p_{\alpha}\ddot{q}_{\alpha}-\frac{dL}{dt}\\
 & =\sum_{\alpha}\frac{d}{dt}\biggparens{\frac{\partial L}{\partial\dot{q}_{\alpha}}}\dot{q}_{\alpha}+\sum_{\alpha}\frac{\partial L}{\partial\dot{q}_{\alpha}}\ddot{q}_{\alpha}-\frac{dL}{dt}\\
 & =\sum_{\alpha}\frac{\partial L}{\partial q_{\alpha}}\dot{q}_{\alpha}+\sum_{\alpha}\frac{\partial L}{\partial\dot{q}_{\alpha}}\ddot{q}_{\alpha}-\frac{dL}{dt}\\
 & =\underbrace{\biggparens{\sum_{\alpha}\frac{\partial L}{\partial q_{\alpha}}\dot{q}_{\alpha}+\sum_{\alpha}\frac{\partial L}{\partial\dot{q}_{\alpha}}\ddot{q}_{\alpha}+\frac{\partial L}{\partial t}}}_{dL/dt}-\frac{dL}{dt}-\frac{\partial L}{\partial t}\\
 & =-\frac{\partial L}{\partial t},
\end{align*}
 from which we conclude that the time derivative of the Hamiltonian
\eqref{eq:DefHamiltonian} is given by 
\begin{equation}
\frac{dH}{dt}=-\frac{\partial L}{\partial t}.\label{eq:TimeDerivHamiltonianFromLagr}
\end{equation}
 The two equalities \eqref{eq:TimeDerivCanonicalMomFromLagr} and
\eqref{eq:TimeDerivHamiltonianFromLagr} look very similar, apart
from an overall minus sign that we will eventually see is not an accident
but has an important physical significance. 

Moreover, we see right away from \eqref{eq:TimeDerivCanonicalMomFromLagr}
that if the Lagrangian is invariant under constant translations along
a specific degree of freedom, $q_{\alpha}\mapsto q_{\alpha}+\parens{\textrm{constant}}$,
so that $\partial L/\partial q_{\alpha}=0$, then the corresponding
canonical momentum $p_{\alpha}$ is conserved, $dp_{\alpha}/dt=0$.
Similarly, we see from \eqref{eq:TimeDerivHamiltonianFromLagr} that
if the Lagrangian is invariant under constant translations in time,
$t\mapsto t+\parens{\textrm{constant}}$, so that $\partial L/\partial t=0$,
then the Hamiltonian $H$ is conserved, $dH/dt=0$. These results
are both special cases of Noether's theorem, which establishes a
general correspondence between continuous symmetries of a classical
system's dynamics and quantities that are conserved when the system
follows its equations of motion.

Taking partial derivatives of the Hamiltonian $H$ with respect to
the canonical variables $q_{\alpha}$ and $p_{\alpha}$, now treated
as independent variables, and regarding $\dot{q}_{\alpha}$ as a function
of the canonical variables, we have 
\begin{align*}
\frac{\partial H}{\partial p_{\alpha}} & =\frac{\partial}{\partial p_{\alpha}}\biggparens{\sum_{\beta}p_{\beta}\dot{q}_{\beta}-L}\\
 & =\dot{q}_{\alpha}+\sum_{\beta}p_{\beta}\frac{\partial\dot{q}_{\beta}}{\partial p_{\alpha}}-\sum_{\beta}\frac{\partial L}{\partial\dot{q}_{\beta}}\frac{\partial\dot{q}_{\beta}}{\partial p_{\alpha}}\\
 & =\dot{q}_{\alpha}+\sum_{\beta}\biggparens{p_{\beta}-\frac{\partial L}{\partial\dot{q}_{\beta}}}\frac{\partial\dot{q}_{\beta}}{\partial p_{\alpha}}\\
 & =\dot{q}_{\alpha}
\end{align*}
 and 
\begin{align*}
\frac{\partial H}{\partial q_{\alpha}} & =\frac{\partial}{\partial q_{\alpha}}\biggparens{\sum_{\beta}p_{\beta}\dot{q}_{\beta}-L}\\
 & =\sum_{\beta}p_{\beta}\frac{\partial\dot{q}_{\beta}}{\partial q_{\alpha}}-\frac{\partial L}{\partial q_{\alpha}}-\sum_{\beta}\frac{\partial L}{\partial\dot{q}_{\beta}}\frac{\partial\dot{q}_{\beta}}{\partial q_{\alpha}}\\
 & =\sum_{\beta}\biggparens{p_{\beta}-\frac{\partial L}{\partial\dot{q}_{\beta}}}\frac{\partial\dot{q}_{\beta}}{\partial q_{\alpha}}-\frac{\partial L}{\partial q_{\alpha}}\\
 & =\frac{d}{dt}\biggparens{\frac{\partial L}{\partial q_{\alpha}}}=\dot{p}_{\alpha},
\end{align*}
 where we have used the Euler-Lagrange equations \eqref{eq:EulerLagrangeEquations}
in the last line. Hence, the Euler-Lagrange equations \eqref{eq:EulerLagrangeEquations}
imply the canonical equations of motion: 
\begin{equation}
\left.\begin{aligned}\dot{q}_{\alpha} & =\frac{\partial H}{\partial p_{\alpha}},\\
\dot{p}_{\alpha} & =-\frac{\partial H}{\partial q_{\alpha}}.
\end{aligned}
\qquad\right\} \label{eq:CanonicalEquationsOfMotion}
\end{equation}

By a similar calculation going the other way, one can also show that
the canonical equations of motion imply the Euler-Lagrange equations,
so the two sets of equations are equivalent. The canonical equations
of motion therefore make it possible to encode the system's dynamics
in an alternative way, known as the Hamiltonian formulation. 

\subsection{The Manifestly Covariant Lagrangian Formulation}

The standard Lagrangian formulation of classical physics treats time
and energy differently from space and momentum, in tension with the
spirit of special relativity. Fortunately, we can recast the Lagrangian
formulation in a more elegant way that puts time and degrees of freedom
on the same footing, with the result that energy and momentum will
naturally also end up on the same footing.\footnote{For an early example of this formalism, see \citep{Dirac:1926rqmacs}.
See also \citep{Dirac:1964loqm}. For more modern reviews, see \citep{DeriglazovRizzuti:2011rifcmse,Souriau:1997sds}.}

To begin, we turn again to the case of a general classical system
with degrees of freedom $q_{\alpha}$, Lagrangian $L\parens{q,\dot{q},t}$,
and action functional \eqref{eq:ClassicalActionFromLagrangian}, 
\[
S\bracks q\defeq\int dt\,L\parens{q,\dot{q},t}.
\]
 We carry out a smooth, strictly monotonic change of integration variable
from $t$ to a new parameter $\lambda$: 
\begin{equation}
t\mapsto t\parens{\lambda}.\label{eq:ReparamMonotonicFunction}
\end{equation}
 Letting dots now denote derivatives with respect to $\lambda$, 
\begin{equation}
\dot{f}\defeq\frac{df}{d\lambda},\label{eq:ReparamInvDotDeriv}
\end{equation}
 we obtain the following differential relationships: 
\begin{equation}
dt=d\lambda\,\dot{t},\qquad\frac{dq_{\alpha}}{dt}=\frac{\dot{q}_{\alpha}}{\dot{t}}.\label{eq:RaparamInvNewDerivs}
\end{equation}
 Our action functional then becomes 
\begin{equation}
S\bracks q\defeq\int d\lambda\,\dot{t}\,L\parens{q,\dot{q}/\dot{t},t}.\label{eq:ReparamInvActionFunctional}
\end{equation}
 This formula for the system's action functional is reparametrization invariant,
meaning that it would maintain its form if we were to carry out any
subsequent smooth, strictly monotonic change of parametrization $\lambda\mapsto\lambda\parens{\lambda^{\prime}}$:
\begin{equation}
S\bracks q\defeq\int d\lambda^{\prime}\,\frac{dt}{d\lambda^{\prime}}\,L\biggparens{q,\frac{dq}{d\lambda^{\prime}}/\frac{dt}{d\lambda^{\prime}},t}.\label{eq:ReparamInvActionFunctionalAfterReparam}
\end{equation}

Reparametrization invariance is an example of a gauge invariance,
meaning a redefinition of the system's degrees of freedom that leaves
all the system's physically observable features unchanged. A gauge
invariance represents a ``redundancy'' in the mathematical description
of a physical system, in the sense that if we were to redefine the
system's degrees of freedom according to a gauge invariance, then
we would obtain a distinct but mathematically equivalent description
of the same system in the same physical state. 

A gauge invariance should be distinguished from a dynamical symmetry,
which consists of transformations that alter the system's physical
state but leave the system's dynamics unchanged. For example, a Newtonian
system of particles could have dynamical symmetries under translations
or rotations in three-dimensional space, both of which would alter
the system's physical state.

We can formally regard the reparametrization-invariant formula \eqref{eq:ReparamInvActionFunctional}
for the action functional as describing a system with an additional
``degree of freedom'' $t$ and a modified Lagrangian 
\begin{equation}
\mathscr{L}\parens{q,\dot{q},t,\dot{t}}\defeq\dot{t}\,L\parens{q,\dot{q}/\dot{t},t}.\label{eq:ReparamInvLagrangian}
\end{equation}
 Notice that 
\begin{align*}
\frac{\partial\mathscr{L}}{\partial\dot{t}} & =\frac{\partial}{\partial\dot{t}}\parens{\dot{t}\,L\parens{q,\dot{q}/\dot{t},t}}\\
 & =L+\sum_{\alpha}\dot{t}\frac{\partial L}{\partial\parens{\dot{q}_{\alpha}/\dot{t}}}\biggparens{-\frac{\dot{q}_{\alpha}}{\dot{t}^{2}}}\\
 & =L-\sum_{\alpha}p_{\alpha}\frac{dq_{\alpha}}{dt}=-H
\end{align*}
 and 
\begin{align*}
\frac{\partial\mathscr{L}}{\partial\dot{q}_{\alpha}} & =\frac{\partial}{\partial\dot{q}_{\alpha}}\parens{\dot{t}\,L\parens{q,\dot{q}/\dot{t},t}}\\
 & =\dot{t}\frac{\partial L}{\partial\parens{\dot{q}_{\alpha}/\dot{t}}}\frac{1}{\dot{t}}=p_{\alpha}.
\end{align*}
 Thus, the system's new canonical momenta \eqref{eq:DefCanonicalMomenta}
conjugate to our original degrees of freedom $q_{\alpha}$ are the
same as before, $\mathscr{P}_{\alpha}=p_{\alpha}$, whereas the system's
canonical momentum $\mathscr{P}_{t}$ conjugate to $t$ is equal to
\emph{minus} the system's original Hamiltonian $H$: 
\begin{equation}
\left.\begin{aligned}\mathscr{P}_{t} & \defeq\frac{\partial\mathscr{L}}{\partial\dot{t}}=-H,\\
\mathscr{P}_{\alpha} & \defeq\frac{\partial\mathscr{L}}{\partial\dot{q}_{\alpha}}=p_{\alpha}.
\end{aligned}
\qquad\right\} \label{eq:ReparamInvCanonicalMomenta}
\end{equation}
 These formulas motivate introducing ``upper-index'' and ``lower-index''
versions of our canonical variables by mimicking the analogous rules
for the components of the four-vectors that are used in special relativity:
\begin{equation}
\left.\begin{aligned}q^{t} & \defeq c\,t, & q_{t} & \defeq-c\,t,\\
q^{\alpha} & \defeq q_{\alpha},\\
p^{t} & \defeq H/c, & p_{t} & \defeq-H/c,\\
p^{\alpha} & \defeq p_{\alpha}.
\end{aligned}
\qquad\right\} \label{eq:DefRaisingLowerIndicesCanonicalVars}
\end{equation}
 To ensure that we are using the same units for $q^{t}$ and $q^{\alpha}$
and also the same units for $p^{t}$ and $p^{\alpha}$, we have introduced
an arbitrary constant $c$ with units of energy divided by momentum.
(The constant $c$ also has units of distance divided by time, or
speed, but not all classical systems possess a notion of distance.)
Note also that we have defined $p_{t}\defeq\mathscr{P}_{t}/c$.

Applying the extremization condition \eqref{eq:VariationActionFunctional}
to the action functional with respect to the new degrees of freedom
$q^{t}$ and $q^{\alpha}$, we obtain a new set of Euler-Lagrange
equations given by 
\begin{equation}
\left.\begin{aligned}\frac{\partial\mathscr{L}}{\partial q^{t}}-\frac{d}{d\lambda}\biggparens{\frac{\partial\mathscr{L}}{\partial\dot{q}^{t}}} & =0,\\
\frac{\partial\mathscr{L}}{\partial q^{\alpha}}-\frac{d}{d\lambda}\biggparens{\frac{\partial\mathscr{L}}{\partial\dot{q}^{\alpha}}} & =0.
\end{aligned}
\qquad\right\} \label{eq:ReparamInvEulerLagrangeEqs}
\end{equation}
Observe that 
\begin{align*}
 & \frac{\partial\mathscr{L}}{\partial q^{\alpha}}-\frac{d}{d\lambda}\biggparens{\frac{\partial\mathscr{L}}{\partial\dot{q}^{\alpha}}}\\
 & =\dot{t}\frac{\partial L}{\partial q_{\alpha}}-\frac{dt}{d\lambda}\frac{d}{dt}\biggparens{\dot{t}\frac{\partial L}{\partial\parens{\dot{q}_{\alpha}/\dot{t}}}\biggparens{\frac{1}{\dot{t}}}}\\
 & =\dot{t}\biggparens{\frac{\partial L}{\partial q_{\alpha}}-\frac{d}{dt}\biggparens{\frac{\partial L}{\partial\parens{dq_{\alpha}/dt}}}},
\end{align*}
 so the Euler-Lagrange equations for the degrees of freedom $q^{\alpha}$
unsurprisingly give us back our original Euler-Lagrange equations
\eqref{eq:EulerLagrangeEquations}, 
\[
\frac{\partial L}{\partial q_{\alpha}}-\frac{d}{dt}\biggparens{\frac{\partial L}{\partial\parens{dq_{\alpha}/dt}}}=0,
\]
 which, as we recall from \eqref{eq:TimeDerivCanonicalMomFromLagr},
can be written more compactly as 
\[
\frac{dp_{\alpha}}{dt}=\frac{\partial L}{\partial q_{\alpha}}.
\]
 Meanwhile, we also have 
\begin{align*}
 & \frac{\partial\mathscr{L}}{\partial q^{t}}-\frac{d}{d\lambda}\biggparens{\frac{\partial\mathscr{L}}{\partial\dot{q}^{t}}}\\
 & =\frac{1}{c}\frac{\partial\mathscr{L}}{\partial t}-\frac{1}{c}\frac{d}{d\lambda}\biggparens{\frac{\partial\mathscr{L}}{\partial\dot{t}}}\\
 & =\frac{1}{c}\dot{t}\frac{\partial L}{\partial t}-\frac{1}{c}\frac{d}{d\lambda}\biggparens{L+\sum_{\alpha}\dot{t}\frac{\partial L}{\partial\parens{\dot{q}_{\alpha}/\dot{t}}}\biggparens{-\frac{\dot{q}_{\alpha}}{\dot{t}^{2}}}}\\
 & =\frac{1}{c}\dot{t}\frac{\partial L}{\partial t}-\frac{1}{c}\dot{t}\frac{d}{dt}\biggparens{L-\sum_{\alpha}p_{\alpha}\frac{\dot{q}_{\alpha}}{\dot{t}}}\\
 & =\frac{1}{c}\dot{t}\biggparens{\frac{\partial L}{\partial t}+\frac{dH}{dt}},
\end{align*}
 so the Euler-Lagrange equation for $q^{t}$ replicates the equation
\eqref{eq:TimeDerivHamiltonianFromLagr} that relates the total time
derivative of the system's original Hamiltonian $H$ to the partial
time derivative of the system's original Lagrangian $L$, 
\[
\frac{dH}{dt}=-\frac{\partial L}{\partial t}.
\]

We can combine these results in terms of the raised-index versions
$p^{t}$ and $p^{\alpha}$ of the canonical momenta defined in \eqref{eq:DefRaisingLowerIndicesCanonicalVars}
as the symmetric-looking equations 
\begin{equation}
\left.\begin{aligned}\frac{dp^{t}}{dt} & =\frac{\partial L}{\partial q_{t}},\\
\frac{dp^{\alpha}}{dt} & =\frac{\partial L}{\partial q_{\alpha}},
\end{aligned}
\qquad\right\} \label{eq:RaparamInvTimeDerivsCanonicalMomenta}
\end{equation}
 or, equivalently, in terms of $\mathscr{L}$ and derivatives with
respect to $\lambda$ as 
\begin{equation}
\left.\begin{aligned}\dot{p}^{t}\defeq\frac{dp^{t}}{d\lambda} & =\frac{\partial\mathscr{L}}{\partial q_{t}},\\
\dot{p}^{\alpha}\defeq\frac{dp^{\alpha}}{d\lambda} & =\frac{\partial\mathscr{L}}{\partial q_{\alpha}}.
\end{aligned}
\qquad\right\} \label{eq:ReparamInvParamDerivsCanonicalMomenta}
\end{equation}
 Furthermore, we can write our action functional \eqref{eq:ReparamInvActionFunctional}
as 
\begin{align*}
S\bracks q & =\int d\lambda\,\dot{t}\,L=\int d\lambda\,\dot{t}\biggparens{\sum_{\alpha}p_{\alpha}\frac{dq_{\alpha}}{dt}-H}\\
 & =\int d\lambda\,\biggparens{\sum_{\alpha}p_{\alpha}\frac{dt}{d\lambda}\frac{dq_{\alpha}}{dt}-\parens{\dot{q}^{t}/c}H}\\
 & =\int d\lambda\,\biggparens{\sum_{\alpha}p_{\alpha}\dot{q}^{\alpha}+\dot{q}^{t}p_{t}}.
\end{align*}
 That is, rather remarkably, we can recast our action functional in
a form that resembles a Lorentz-invariant dot product, despite the
fact that we have not assumed that our system has anything to do with
special relativity or four-dimensional spacetime: 
\begin{equation}
S\bracks q=\int d\lambda\,\mathscr{L}=\int d\lambda\,\bigparens{p_{t}\dot{q}^{t}+\sum_{\alpha}p_{\alpha}\dot{q}^{\alpha}}.\label{eq:ReparamActionFunctionalFromCanonicalMomenta}
\end{equation}
 We therefore refer to this framework as the manifestly covariant Lagrangian formulation
for our classical system.

Introducing a square matrix $\eta\defeq\mathrm{diag}\parens{-1,1,\dotsc}$
that naturally generalizes the Minkowski metric tensor from special
relativity, 
\begin{equation}
\eta\defeq\begin{pmatrix}-1 & 0 & 0\\
0 & 1 & 0\\
0 & 0 & \smash{{\ddots}}
\end{pmatrix},\label{eq:ReparamGeneralizedMinkMetric}
\end{equation}
 we can write the system's action functional \eqref{eq:ReparamActionFunctionalFromCanonicalMomenta}
in matrix form as 
\begin{equation}
S\bracks q=\int d\lambda\,\begin{pmatrix}p^{t} & p^{\alpha}\end{pmatrix}\,\eta\,\begin{pmatrix}\dot{q}^{t}\\
\dot{q}^{\alpha}
\end{pmatrix},\label{eq:ReparamMatrixFormActionFunctional}
\end{equation}
 where $p^{\alpha}$ and $\dot{q}^{\alpha}$ here are notational abbreviations
for their whole lists indexed by $\alpha$. This expression for $S\bracks q$
immediately suggests the consideration of systems whose action functionals
have a symmetry under rigid linear transformations of the form 
\begin{equation}
\begin{pmatrix}q^{t}\\
q^{\alpha}
\end{pmatrix}\mapsto\Lambda\begin{pmatrix}q^{t}\\
q^{\alpha}
\end{pmatrix},\quad\begin{pmatrix}p^{t}\\
p^{\alpha}
\end{pmatrix}\mapsto\Lambda\begin{pmatrix}p^{t}\\
p^{\alpha}
\end{pmatrix}\label{eq:ReparamGeneralizedLorentzTransform}
\end{equation}
 for constant matrices $\Lambda$ that preserve the generalized Minkowski
metric tensor $\eta$ in the sense that 
\begin{equation}
\Lambda^{\transp}\eta\Lambda=\eta.\label{eq:ReparamGeneralizedMinkMetricInv}
\end{equation}
 The matrices $\Lambda$ therefore represent generalizations of Lorentz
transformations.

Recall the group $O\parens N$ of orthogonal $N\times N$ matrices
$R$, meaning matrices that preserve the $N\times N$ identity matrix
$1\defeq\mathrm{diag}\parens{1,1,\dotsc}$, 
\begin{equation}
R^{\transp}R=R^{\transp}1R=1.\label{eq:RotationPreserveIdentitymatrix}
\end{equation}
 Letting $N$ denote the system's original number of degrees of freedom
$q_{\alpha}$, we see that the set of generalized Lorentz-transformation
matrices $\Lambda$ preserve the $\parens{N+1}\times\parens{N+1}$
matrix $\eta\defeq\mathrm{diag}\parens{-1,1,\dotsc}$, so we correspondingly
refer to them as making up the group $O\parens{1,N}$.

The formula \eqref{eq:ReparamActionFunctionalFromCanonicalMomenta}
for the action functional also implies that the new ``Hamiltonian''
$\mathscr{H}$, defined in line with \eqref{eq:DefHamiltonian}, trivially
vanishes, and therefore (at least classically) does not hold any physical
meaning: 
\begin{equation}
\mathscr{H}\defeq p_{t}\dot{q}^{t}+\sum_{\alpha}p_{\alpha}q^{\alpha}-\mathscr{L}=0.\label{eq:ReparamHamiltonianVanishes}
\end{equation}
 This equation is closely related to the fact that arbitrary changes
of parametrization represent a gauge invariance of the system and
likewise do not have any physical meaning.

\section{Spacetime in Special Relativity}

We now turn to a brief review of special relativity.\footnote{For a more extensive introduction, see the opening chapters of \citep{Schutz:2009fcgr}.}

\subsection{Spacetime and Four-Vectors}

In special relativity, time $t$ and space $\vec x\defeq\parens{x,y,z}$
join together to form four-dimensional spacetime coordinates, 
\begin{align}
x^{\mu} & \defeq\parens{x^{t},x^{x},x^{y},x^{z}}^{\mu}\nonumber \\
 & \qquad\defeq\parens{c\,t,\vec x}^{\mu}\defeq\parens{c\,t,x,y,z}^{\mu},\label{eq:4DSpacetimeCoordinates}
\end{align}
 where $c$ is the speed of light. We will use Greek letters $\alpha,\beta,\dots,\mu,\nu,\dotsc$
for Lorentz indices, which will each run through the four possible
values $t,x,y,z$, and we will use Latin indices $i,j,k,\dotsc$ for
the spatial values $x,y,z$, where we will consistently employ Cartesian
coordinate systems.

Defining the (3+1)-dimensional Minkowski metric tensor by 
\begin{equation}
\eta_{\mu\nu}\defeq\eta^{\mu\nu}\defeq\begin{pmatrix}-1 & 0 & 0 & 0\\
0 & 1 & 0 & 0\\
0 & 0 & 1 & 0\\
0 & 0 & 0 & 1
\end{pmatrix}_{\mathclap{\mu\nu}},\label{eq:DefMinkMetricTensor}
\end{equation}
 and employing Einstein summation notation, we can raise and lower
indices on the components of four-vectors according to $v_{\mu}\defeq\eta_{\mu\nu}v^{\nu}$
and $w^{\mu}\defeq\eta^{\mu\nu}w_{\nu}$, with the following results:
\begin{equation}
\left.\begin{aligned}v^{t} & =-v_{t},\\
v^{x} & =\ \ v_{x},\\
v^{y} & =\ \ v_{y},\\
v^{z} & =\ \ v_{z}.
\end{aligned}
\quad\right\} \label{eq:RaiseLowerIndices4Vec}
\end{equation}
 We let $\tud{\Lambda}{\mu}{\nu}$ be a $4\times4$ Lorentz-transformation
matrix, meaning that $\tud{\Lambda}{\mu}{\nu}$ is an element of $O\parens{1,3}$
and therefore preserves the Minkowski metric tensor $\eta_{\mu\nu}$
in the sense that 
\begin{equation}
\tud{\Lambda}{\mu}{\rho}\eta_{\mu\nu}\tud{\Lambda}{\nu}{\sigma}=\eta_{\rho\sigma},\label{eq:LorentzTransfsPreserveMinkMetricIndices}
\end{equation}
 or, in matrix notation, 
\begin{equation}
\Lambda^{\transp}\eta\Lambda=\eta.\label{eq:LorentzTransfsPreserveMinkMetric}
\end{equation}
 Then Lorentz transformations of four-vectors $v^{\mu}$, meaning
linear transformations of the form 
\begin{equation}
v^{\mu}\mapsto\tud{\Lambda}{\mu}{\nu}v^{\nu},\label{eq:Def4DLorentzTransf4Vec}
\end{equation}
 preserve four-dimensional dot products defined by 
\begin{equation}
v\dotprod w\defeq v_{\nu}w^{\nu}=\eta_{\mu\nu}v^{\mu}w^{\nu}.\label{eq:LorentzInvDotProd}
\end{equation}

Four-vectors $v^{\mu}$ are classified as timelike, null, or spacelike
according to whether the dot product of $v^{\mu}$ with itself is
respectively negative, zero, or positive: 
\begin{equation}
v^{2}\defeq v\dotprod v\ \begin{cases}
<0 & \textrm{timelike},\\
=0 & \textrm{null},\\
>0 & \textrm{spacelike}.
\end{cases}\label{eq:DefTimelikeNulLSpacelike}
\end{equation}
 The Lorentz invariance of the dot product \eqref{eq:LorentzInvDotProd}
ensures that this classification is invariant and therefore well-defined
under Lorentz transformations.

\subsection{The Spacetime Transformation Groups}

The collection $O\parens{1,3}$ of all possible Lorentz transformations
\eqref{eq:Def4DLorentzTransf4Vec}, 
\[
v^{\mu}\mapsto\tud{\Lambda}{\mu}{\nu}v^{\nu},
\]
 is called the Lorentz group.\footnote{For a comprehensive presentation of the group theory underlying special
relativity, see \citep{Weinberg:1996tqtfi}.} The largest subgroup that excludes parity transformations, 
\begin{equation}
\Lambda_{\textrm{parity}}=\mathrm{diag}\parens{1,-1,-1,-1}=\begin{pmatrix}1 & 0 & 0 & 0\\
0 & -1 & 0 & 0\\
0 & 0 & -1 & 0\\
0 & 0 & 0 & -1
\end{pmatrix},\label{eq:ParityTransf}
\end{equation}
 is called the proper Lorentz group and is denoted by $SO\parens{1,3}$,
mirroring the notation $SO\parens N$ for $N\times N$ rotation matrices
$R$ that do not involve parity transformations. The largest subgroup
of the Lorentz group that excludes time-reversal transformations,
\begin{equation}
\Lambda_{\textrm{time-reversal}}=\mathrm{diag}\parens{-1,1,1,1}=\begin{pmatrix}-1 & 0 & 0 & 0\\
0 & 1 & 0 & 0\\
0 & 0 & 1 & 0\\
0 & 0 & 0 & 1
\end{pmatrix},\label{eq:TimeReversalTransf}
\end{equation}
 is called the orthochronous Lorentz group and is denoted by $O^{+}\parens{1,3}$
or $O^{\uparrow}\parens{1,3}$. The set of all Lorentz transformations
that can be reduced smoothly to the identity transformation $\Lambda=1$
cannot include parity or time-reversal transformations, and is called
the proper orthochronous Lorentz group $SO^{+}\parens{1,3}$ or $SO^{\uparrow}\parens{1,3}$.

A simple calculation shows that for four-vectors $v^{\mu}$ that are
timelike or null, the \emph{sign} of the temporal component $v^{t}$
is invariant under orthochronous Lorentz transformations $v^{\mu}\mapsto\tud{\Lambda}{\mu}{\nu}v^{\nu}$:
\begin{equation}
v^{2}\leq0\implies\textrm{sign of \ensuremath{v^{t}} is invariant under \ensuremath{O^{+}\parens{1,3}}.}\label{eq:CausalPropertySignTemporalComponentTimelikeNull4Vecs}
\end{equation}
As a consequence, future-directed ($v^{t}>0$) four-vectors that are
timelike or null remain future-directed under orthochronous Lorentz
transformations, with a similar statement for past-directed ($v^{t}<0$)
four-vectors that are timelike or null. These properties ensure that
if the displacement between two spacetime points is timelike or null,
then their chronological ordering is an invariant fact of nature.
By contrast, the temporal components $v^{t}$ of spacelike four-vectors
($v^{2}>0$) can change sign under orthochronous Lorentz transformations,
a behavior that is closely related to the breakdown of simultaneity
in special relativity.

We can also consider additive shifts in the four-dimensional coordinates
\eqref{eq:4DSpacetimeCoordinates} by constants $a^{\mu}$: 
\begin{equation}
x^{\mu}\mapsto x^{\mu}+a^{\mu}.\label{eq:4DTranslCoords}
\end{equation}
These transformations make up the spacetime-translation group, which
is isomorphic to $\mathbb{R}^{4}$ but is denoted by $\mathbb{R}^{1,3}$
to emphasize the mathematical and physical distinctions between time
and space.

Combining spacetime translations with Lorentz transformations of the
spacetime coordinates $x^{\mu}$ gives the Poincaré group: 
\begin{equation}
x^{\mu}\mapsto\tud{\Lambda}{\mu}{\nu}x^{\nu}+a^{\mu}.\label{eq:4DPoincareTransfCoords}
\end{equation}
 Like the Lorentz group, the Poincaré group has proper and orthochronous
subgroups that are respectively defined by dropping all Lorentz transformations
that involve parity or time-reversal transformations.\footnote{As a mathematical aside, the Poincaré group is formally denoted by
the semi-direct product $\mathbb{R}^{1,3}\rtimes O\parens{1,3}$,
which generalizes the notion of a direct product $G=H_{1}\times H_{2}$
to the case in which the second factor $H_{2}$ is not necessarily
a normal subgroup of the overall group $G$.}

\section{Transitive Group Actions of the Poincaré Group}

The set of all physical transformations $\parens{q,p}\mapsto\parens{q^{\prime},p^{\prime}}$
that can be carried out on a system's state $\parens{q,p}$ in its
phase space are collectively called a group action on the system's
phase space. If we include translations in time among these physical
transformations, then by starting with a single convenient choice
of reference state $\parens{q_{\refvalue},p_{\refvalue}}$, we can
reach every other possible state that the system can occupy. The group
action provided by the system's phase space is therefore ``irreducible,''
or, more precisely, transitive, referring to the fact that no proper
subset of the system's phase space can be dropped without violating
the group action. One then says that the phase space serves as a homogeneous
space for the group of physical transformations.

As we will show, the different possible transitive group actions (or
homogeneous spaces) of the Poincaré group turn out to provide a complete
classification of the phase spaces of the different categories of
particles in physics, in parallel with Wigner's method for classifying
quantum particle-types by identifying their Hilbert spaces as irreducible
representations of the Poincaré group.\footnote{For alternative classical approaches to this classification problem,
see \citep{SudarshanMukunda:1974cdamp,BalachandranMarmoSkagerstamStern:1983gsfb,Souriau:1997sds,Rivas:2002ktsp}.}

\subsection{Systems Singled Out by the Poincaré Group}

To start, we note that the Poincaré group \eqref{eq:4DPoincareTransfCoords}
naturally singles out classical systems that have three physical degrees
of freedom $\parens{q_{x},q_{y},q_{z}}=\vec X\defeq\parens{X,Y,Z}$
and therefore three corresponding canonical momenta $\vec p=\parens{p_{x},p_{y},p_{z}}$.
It follows that the system's manifestly covariant Lagrangian formulation
involves four spacetime degrees of freedom 
\begin{align}
X^{\mu} & \defeq\parens{q^{t},q^{x},q^{y},q^{z}}^{\mu}\nonumber \\
 & =\parens{c\,T,X,Y,Z}^{\mu}\defeq\parens{c\,T,\vec X}^{\mu},\label{eq:PoincareActionSystemCoords}
\end{align}
 together with a canonical four-momentum 
\begin{align}
p^{\mu} & \defeq\parens{p^{t},p^{x},p^{y},p^{z}}^{\mu}\nonumber \\
 & \defeq\parens{E/c,\vec p}^{\mu}\label{eq:PoincareActionSystem4Momentum}
\end{align}
 whose individual components, in lower-index form $p_{\mu}$, are
defined in terms of the system's covariant Lagrangian $\mathscr{L}$
in accordance with \eqref{eq:ReparamInvCanonicalMomenta}, 
\begin{equation}
p_{\mu}\defeq\frac{\partial\mathscr{L}}{\partial\dot{X}^{\mu}}.\label{eq:PoincareActionSystem4MomFromLagrangian}
\end{equation}
 Here dots denote derivatives with respect to the arbitrary parameter
$\lambda$, 
\begin{equation}
\dot{X}^{\mu}\defeq\frac{dX^{\mu}}{d\lambda},\label{eq:DotCoordParamDeriv}
\end{equation}
 we have identified the system's energy $E$ as 
\begin{equation}
E\defeq H\defeq p^{t}c,\label{eq:EnergyAsHamiltonianAsTemporalMom}
\end{equation}
 and candidate trajectories of the system are now called worldlines.

\subsection{Angular Momentum and Spin}

In analogy with the Newtonian definition $\vec L\defeq\vec X\crossprod\vec p$
of an object's orbital angular momentum, whose individual components
are 
\begin{align}
L_{k} & =X_{i}p_{j}-X_{j}p_{i},\nonumber \\
 & \quad\textrm{with \ensuremath{\parens{i,j,k}=\parens{x,y,z}}, \ensuremath{\parens{z,x,y}}, or \ensuremath{\parens{y,z,x}}},\label{eq:3DOrbAngMomComps}
\end{align}
we will find it convenient to introduce an antisymmetric tensor 
\begin{equation}
L^{\mu\nu}\defeq X^{\mu}p^{\nu}-X^{\nu}p^{\mu}=-L^{\nu\mu}\label{eq:Def4DOrbitalAngularMomentum}
\end{equation}
 whose spatial components $L^{ij}$ (that is, for $i,j$ each taking
the values $x,y,z$) encode the components of $\vec L$. We will accordingly
refer to $L^{\mu\nu}$ as the system's orbital angular-momentum tensor,
although one should keep in mind that its temporal components $L^{ti}$
(for $i$ a spatial index) are not angular momenta. Indeed, if the
system's energy \eqref{eq:EnergyAsHamiltonianAsTemporalMom} is nonzero,
$E\defeq p^{t}c\ne0$, then we can write these temporal components
as 
\begin{align}
L^{ti} & =X^{t}p^{i}-X^{i}p^{t}=c\,T\,p^{i}-X^{i}E/c\nonumber \\
 & =-\frac{E}{c}\biggparens{X^{i}-\frac{p^{i}c^{2}}{E}T}.\label{eq:OrbAngMomTemporalCompsAsVelFormula}
\end{align}
 We will see later that the factor $\vec pc^{2}/E$, which has units
of distance divided by time, will typically yield the system's three-dimensional
physical propagation velocity $\vec v\defeq d\vec X/dt$ through space,
so the quantity in parentheses will turn out to be related to the
system's \emph{linear} motion. 

To be as general as possible, we can also allow the system to possess
an \emph{intrinsic} notion of angular momentum, called spin, that
does not involve the system's spacetime coordinates $X^{\mu}$ or
its four-momentum $p^{\mu}$, and that can be encoded in an antisymmetric
tensor 
\begin{equation}
S^{\mu\nu}=-S^{\nu\mu},\label{eq:Def4DSpinTensor}
\end{equation}
 called the system's spin tensor. The system's \emph{total} angular
momentum is then represented by an antisymmetric tensor defined as
the sum of the tensors representing the orbital and spin contributions:
\begin{equation}
J^{\mu\nu}\defeq L^{\mu\nu}+S^{\mu\nu}=-J^{\nu\mu}.\label{eq:Def4DTotalAngMomTensor}
\end{equation}
 We will refer to $J^{\mu\nu}$ as the system's total angular-momentum tensor.

We can define the following three-vectors from the independent components
of $J^{\mu\nu}$ and $S^{\mu\nu}$: 
\begin{align}
\vec J & \defeq\parens{J_{x},\,\,\,J_{y},\,\,J_{z}}\defeq\parens{J^{yz},J^{zx},J^{xy}},\label{eq:DefTotAngMom3Vec}\\
\vec K & \defeq\parens{K_{x},K_{y},K_{z}}\defeq\parens{J^{tx},J^{ty},J^{tz}},\label{eq:DefTotBoost3Vec}\\
\vec S & \defeq\parens{S_{x},\,\,S_{y},\,\,S_{z}}\defeq\parens{S^{yz},S^{zx},S^{xy}},\label{eq:DefSpinAngMom3Vec}\\
\tilde{\vec S} & \defeq\parens{\tilde{S}_{x},\,\,\tilde{S}_{y},\,\,\tilde{S}_{z}}\defeq\parens{S^{tx},S^{ty},S^{tz}}.\label{eq:DefDualSpinBoost3Vec}
\end{align}
 We will call $\vec S$ the system's spin three-vector and $\tilde{\vec S}$
its dual spin-three vector.

We can now write the system's total angular-momentum tensor $J^{\mu\nu}$
and its spin tensor $S^{\mu\nu}$ as 
\begin{equation}
J^{\mu\nu}\defeq\begin{pmatrix}0 & K_{x} & K_{y} & K_{z}\\
-K_{x} & 0 & J_{z} & -J_{y}\\
-K_{y} & -J_{z} & 0 & J_{x}\\
-K_{z} & J_{y} & -J_{x} & 0
\end{pmatrix}^{\mathclap{\mu\nu}},\label{eq:TotAngMomTensorFrom3Vecs}
\end{equation}
\begin{equation}
S^{\mu\nu}\defeq\begin{pmatrix}0 & \tilde{S}_{x} & \tilde{S}_{y} & \tilde{S}_{z}\\
-\tilde{S}_{x} & 0 & S_{z} & -S_{y}\\
-\tilde{S}_{y} & -S_{z} & 0 & S_{x}\\
-\tilde{S}_{z} & S_{y} & -S_{x} & 0
\end{pmatrix}^{\mathclap{\mu\nu}}.\label{eq:SpinTensorFrom3Vecs}
\end{equation}
Note that if $\vec S=0$, then $\vec J=\vec L=\vec X\crossprod\vec p$
reduces to the usual Newtonian definition \eqref{eq:3DOrbAngMomComps}
of orbital angular momentum. 

\subsection{Defining a System by a Transitive Group Action of the Poincaré Group}

The state of our system in its phase space is fully determined by
knowing the values of the system's spacetime coordinates $X^{\mu}$,
its four-momentum $p^{\mu}$, and its spin tensor $S^{\mu\nu}$, which
together determine the orbital angular-momentum tensor $L^{\mu\nu}$
and the total angular-momentum tensor $J^{\mu\nu}$. We can therefore
define a transitive group action of the Poincaré group on the system's
phase space by defining what Poincaré transformations do to the values
of $X^{\mu}$, $p^{\mu}$, and $S^{\mu\nu}$ that define the system's
state $\parens{X,p,S}$.

Specifically, we define the action of Lorentz transformations on the
system's state $\parens{X,p,S}$ by generalizing the transformation
rule \eqref{eq:Def4DLorentzTransf4Vec} to the statement that every
free upper Lorentz index on $X^{\mu}$, $p^{\mu}$, and $S^{\mu\nu}$
receives a linear factor of a shared Lorentz-transformation matrix
$\Lambda$: 
\begin{align}
X^{\mu} & \mapsto\tud{\Lambda}{\mu}{\nu}X^{\nu},\label{eq:Def4DLorentzTransfCoords}\\
p^{\mu} & \mapsto\tud{\Lambda}{\mu}{\nu}p^{\nu},\label{eq:Def4DLorentzTransf4Mom}\\
S^{\mu\nu} & \mapsto\tud{\Lambda}{\mu}{\rho}\tud{\Lambda}{\nu}{\sigma}S^{\rho\sigma}=\tud{\Lambda}{\mu}{\rho}S^{\rho\sigma}\tdu{\parens{\Lambda^{\transp}}}{\sigma}{\nu}.\label{eq:Def4DLorentzTransfSpinTensor}
\end{align}
 It follows from the definitions \eqref{eq:Def4DOrbitalAngularMomentum}
of $L^{\mu\nu}$ and \eqref{eq:Def4DTotalAngMomTensor} of $J^{\mu\nu}$
that we have the additional Lorentz-transformation rules 
\begin{align}
L^{\mu\nu} & \mapsto\tud{\Lambda}{\mu}{\rho}\tud{\Lambda}{\nu}{\sigma}L^{\rho\sigma}=\tud{\Lambda}{\mu}{\rho}L^{\rho\sigma}\tdu{\parens{\Lambda^{\transp}}}{\sigma}{\nu},\label{eq:4DLorentzTransfOrbAngMom}\\
J^{\mu\nu} & \mapsto\tud{\Lambda}{\mu}{\rho}\tud{\Lambda}{\nu}{\sigma}J^{\rho\sigma}=\tud{\Lambda}{\mu}{\rho}J^{\rho\sigma}\tdu{\parens{\Lambda^{\transp}}}{\sigma}{\nu}.\label{eq:4DLorentzTransfTotalAngMom}
\end{align}
 Meanwhile, we define the action of spacetime translations on the
system's state $\parens{X,p,S}$ solely as \eqref{eq:4DTranslCoords}
for the spacetime coordinates $X^{\mu}$, with the system's four-momentum
$p^{\mu}$ and spin tensor $S^{\mu\nu}$ unchanged: 
\begin{align}
X^{\mu} & \mapsto X^{\mu}+a^{\mu},\label{eq:Def4DConstTranslCoords}\\
p^{\mu} & \mapsto p^{\mu},\label{eq:Def4DConstTransl4Mom}\\
S^{\mu\nu} & \mapsto S^{\mu\nu}.\label{eq:Def4DConstTranslSpinTensor}
\end{align}
 These definitions then determine the additional translation rules
\begin{align}
L^{\mu\nu} & \mapsto L^{\mu\nu}+a^{\mu}p^{\nu}-a^{\nu}p^{\mu},\label{eq:4DConstTranslOrbAngMom}\\
J^{\mu} & \mapsto J^{\mu\nu}+a^{\mu}p^{\nu}-a^{\nu}p^{\mu}.\label{eq:4DConstTranslTotalAngMom}
\end{align}
 We can then construct general Poincaré transformations from combinations
of Lorentz transformations and spacetime translations.

One can check that the three-vectors $\vec J$, $\vec K$, $\vec S$,
and $\tilde{\vec S}$ defined in \eqref{eq:DefTotAngMom3Vec}\textendash \eqref{eq:DefDualSpinBoost3Vec}
all indeed transform as three-vectors under proper rotations. One
can also show that $\vec K$ and $\tilde{\vec S}$ transform as proper
vectors (or polar vectors) under parity transformations \eqref{eq:ParityTransf},
\begin{equation}
\left.\begin{aligned}\vec K & \mapsto-\vec K,\\
\tilde{\vec S} & \mapsto-\tilde{\vec S},
\end{aligned}
\quad\right\} \ \parens{\textrm{parity}}\label{eq:Boost3VecsParityTransf}
\end{equation}
 whereas $\vec J$ and $\vec S$ are pseudovectors (or axial vectors),
meaning that they do not change sign under parity transformations:
\begin{equation}
\left.\begin{aligned}\vec J & \mapsto\vec J,\\
\vec S & \mapsto\vec S.
\end{aligned}
\quad\right\} \ \parens{\textrm{parity}}\label{eq:AngMom3VecsParityTransf}
\end{equation}

If the system's phase space provides a transitive group action of
the Poincaré group, then, by construction, every state $\parens{X,p,S}$
can be reached by starting with an arbitrary choice of reference state
\begin{equation}
\parens{X_{\refvalue},p_{\refvalue},S_{\refvalue}}\label{eq:DefPhaseSpacePointReferenceState}
\end{equation}
 and then acting on it with an appropriate choice of Poincaré transformation
$\parens{a,\Lambda}$: 
\begin{equation}
\parens{X,p,S}\defeq\parens{\Lambda X_{\refvalue}+a,\Lambda p_{\refvalue},\Lambda S_{\refvalue}\Lambda^{\transp}}.\label{eq:PhaseSpacePointFromPoincTransfFromRef}
\end{equation}
 That is, 
\begin{align}
X & \defeq\Lambda X_{\refvalue}+a,\label{eq:SpacetimeCoordFromRef}\\
p & \defeq\Lambda p_{\refvalue},\label{eq:FourMomentumFromRef}\\
S & \defeq\Lambda S_{\refvalue}\Lambda^{\transp},\label{eq:SpinTensorFromRef}
\end{align}
 or, displaying indices explicitly, 
\begin{align}
X^{\mu} & \defeq\tud{\Lambda}{\mu}{\nu}X_{\refvalue}^{\nu}+a^{\mu},\label{eq:SpacetimeCoordFromRefIndices}\\
p^{\mu} & \defeq\tud{\Lambda}{\mu}{\nu}p_{\refvalue}^{\nu},\label{eq:FourMomentumFromRefIndices}\\
S^{\mu\nu} & \defeq\tud{\Lambda}{\mu}{\rho}S_{\refvalue}^{\rho\sigma}\tdu{\parens{\Lambda^{\transp}}}{\sigma}{\nu}.\label{eq:SpinTensorFromRefIndices}
\end{align}

Without loss of generality, we will always take the reference value
of the system's spacetime point to be at the origin: 
\begin{equation}
X_{\refvalue}^{\mu}\defeq0.\label{eq:Def4DRefCoordsAsOrigin}
\end{equation}
 Due to the transformation rule \eqref{eq:SpacetimeCoordFromRefIndices},
the system's spacetime point $X^{\mu}$ in any other state $\parens{X,p,S}$
can then be identified with the translation-group four-vector $a^{\mu}$,
so we will refer to $a^{\mu}$ as $X^{\mu}$ in our work ahead, 
\begin{equation}
X^{\mu}\defeq a^{\mu},\label{eq:Def4DCoordsAsTranslParams}
\end{equation}
 keeping in mind that these variables are independent of the Lorentz-transformation
matrix $\tud{\Lambda}{\mu}{\nu}$. We will choose the reference values
$p_{0}^{\mu}$ and $S_{0}^{\mu\nu}$ in \eqref{eq:DefPhaseSpacePointReferenceState}
on a case-by-case basis later.

\subsection{The Pauli-Lubanski Pseudovector}

Introducing the totally antisymmetric, four-index Levi-Civita symbol,
\begin{align}
\epsilon_{\mu\nu\rho\sigma} & \defeq\begin{cases}
+1 & \textrm{for \ensuremath{\mu\nu\rho\sigma} an even permutation of \ensuremath{txyz}},\\
-1 & \textrm{for \ensuremath{\mu\nu\rho\sigma} an odd permutation of \ensuremath{txyz}},\\
0 & \textrm{otherwise}
\end{cases}\nonumber \\
 & =-\epsilon^{\mu\nu\rho\sigma},\label{eq:4DLeviCivita}
\end{align}
  we can form a convenient mathematical object, called the Pauli-Lubanski pseudovector
$W^{\mu}$, by contracting the Lorentz indices of the system's four-momentum
$p^{\mu}$ and the total angular-momentum tensor $J^{\mu\nu}$ with
the indices of $\epsilon^{\mu\nu\rho\sigma}$:\footnote{The minus sign in this definition is a reflection of our metric sign
conventions.} 
\begin{equation}
W^{\mu}\defeq-\frac{1}{2}\epsilon^{\mu\nu\rho\sigma}p_{\nu}J_{\rho\sigma}.\label{eq:DefPauliLubanski4Vec}
\end{equation}

Decomposing the total angular-momentum tensor as in \eqref{eq:Def4DTotalAngMomTensor}
into its orbital \eqref{eq:Def4DOrbitalAngularMomentum} and spin
\eqref{eq:Def4DSpinTensor} contributions, 
\begin{align*}
J_{\rho\sigma} & =L_{\rho\sigma}+S_{\rho\sigma}\\
 & =X_{\rho}p_{\sigma}-X_{\sigma}p_{\rho}+S_{\rho\sigma},
\end{align*}
 the contributions from the orbital-angular momentum tensor $L_{\rho\sigma}$
cancel out of the definition of $W^{\mu}$, so we can replace the
total angular-momentum tensor $J_{\rho\sigma}$ with just its spin
contribution $S_{\rho\sigma}$ in the formula for $W^{\mu}$: 
\begin{equation}
W^{\mu}=-\frac{1}{2}\epsilon^{\mu\nu\rho\sigma}p_{\nu}S_{\rho\sigma}.\label{eq:PauliLubanski4VecFromSpin}
\end{equation}
 It follows from a straightforward calculation that we can express
the Pauli-Lubanski pseudovector in terms of the spin three-vector
$\vec S$ defined in \eqref{eq:DefSpinAngMom3Vec}, the dual spin
three-vector $\tilde{\vec S}$ defined in \eqref{eq:DefDualSpinBoost3Vec},
and the components of the system's four-momentum $p^{\mu}=\parens{E/c,\vec p}^{\mu}$
as 
\begin{equation}
W^{\mu}=\parens{\vec p\dotprod\vec S,\ \parens{E/c}\vec S-\vec p\crossprod\tilde{\vec S}}^{\mu}.\label{eq:PauliLubanskiFrom3Vecs}
\end{equation}

The formula \eqref{eq:PauliLubanski4VecFromSpin} makes manifest that
the Pauli-Lubanski pseudovector does not involve the spacetime coordinates
$X^{\mu}$, so it is invariant under translation transformations \eqref{eq:Def4DConstTranslCoords}\textendash \eqref{eq:4DConstTranslTotalAngMom}:
\begin{equation}
W^{\mu}\mapsto W^{\mu}\quad\parens{\textrm{spacetime translations}}.\label{eq:PauliLubanskiTranslInvariant}
\end{equation}
  Meanwhile, under Lorentz transformations of $p_{\nu}$ and $S_{\rho\sigma}$,
$W^{\mu}$ transforms as 
\begin{equation}
W^{\mu}\mapsto\det\parens{\Lambda}\tud{\Lambda}{\mu}{\nu}W^{\nu},\label{eq:PauliLubanskiLorentzTransf}
\end{equation}
 where $\det\parens{\Lambda}$ is the determinant of $\tud{\Lambda}{\mu}{\nu}$.
Hence, under parity transformations $\Lambda_{\textrm{parity}}$,
for which $\det\parens{\Lambda_{\textrm{parity}}}=-1$, $W^{\mu}$
transforms oppositely to the way that ordinary four-vectors transform:
\begin{equation}
W^{t}\mapsto-W^{t},\quad W^{i}=W^{i}\quad\parens{\textrm{parity}}.\label{eq:PauliLubanskiParityTransf}
\end{equation}
 It is because of this transformation behavior that $W^{\mu}$ is
called a pseudovector. 

\subsection{Invariant Quantities of a Transitive Group Action of the Poincaré
Group}

Notice that the quantities $p^{2}\defeq p_{\mu}p^{\mu}$, $W^{2}\defeq W_{\mu}W^{\mu}$,
and $S^{2}\defeq S_{\mu\nu}S^{\mu\nu}$ are invariant under Poincaré
transformations, meaning that they are invariant under all Lorentz
transformations (whether or not parity and time-reversal transformations
are involved) as well as under all spacetime translations. These quantities
therefore each have a single, constant value for all states in any
phase space that constitutes a transitive group action of the Poincaré
group, and so, in particular, have constant values along the system's
worldline.\footnote{Quantities that have fixed values in a transitive group action or
in an irreducible representation of a given transformation group are
formally called Casimir invariants.}

We name these invariant quantities according to 
\begin{align}
p^{2} & \defeq p_{\mu}p^{\mu}\defeq-m^{2}c^{2},\label{eq:Def4DMassSquaredAsInvariant}\\
W^{2} & \defeq W_{\mu}W^{\mu}\defeq w^{2},\label{eq:SquarePauliLubanskiAsInvariant}\\
\frac{1}{2}S^{2} & \defeq\frac{1}{2}S_{\mu\nu}S^{\mu\nu}\defeq s^{2}.\label{eq:Def4DSpinSquaredAsInvariant}
\end{align}
 The scalar constant $m$ has units of momentum-squared divided by
energy (that is, units of mass), the scalar constant $w$ has units
of momentum multiplied by energy multiplied by time, and the scalar
constant $s$ has units of energy multiplied by time (that is, units
of angular momentum).  

Note that $w^{2}$ and $s^{2}$ having fixed values does not imply
any sort of quantization, any more than $m^{2}$ being fixed implies
quantization. In our classical context, we are essentially working
in the limit of large quantum numbers in which $w^{2}$ and $s^{2}$
are invariant but are otherwise permitted to take on any one of a
continuous range of possible real values.

In terms of the spin three-vector $\vec S$ defined in \eqref{eq:DefSpinAngMom3Vec}
and the dual spin three-vector $\tilde{\vec S}$ defined in \eqref{eq:DefDualSpinBoost3Vec},
we can write the invariant quantity $s^{2}$ as 
\begin{equation}
s^{2}\defeq\frac{1}{2}S_{\mu\nu}S^{\mu\nu}=\vec S^{2}-\tilde{\vec S}^{2}.\label{eq:4DSpinSquaredInvariantFrom3Vecs}
\end{equation}
 We can also contract two copies of the spin tensor $S^{\mu\nu}$
with the Levi-Civita symbol \eqref{eq:4DLeviCivita} to obtain another
quantity with the same units as $s^{2}$: 
\begin{equation}
\tilde{s}^{2}\defeq\frac{1}{8}\epsilon_{\mu\nu\rho\sigma}S^{\mu\nu}S^{\rho\sigma}=\vec S\dotprod\tilde{\vec S}.\label{eq:Def4DDualSpinSquaredInvariant}
\end{equation}
 This quantity is invariant under spacetime translations and also
under \emph{proper orthochronous} Lorentz transformations. However,
due to the transformation rules \eqref{eq:Boost3VecsParityTransf}
and \eqref{eq:AngMom3VecsParityTransf}, $\tilde{s}^{2}$ changes
by an overall sign under parity transformations, so it is called
a pseudoscalar. 

As was true for the scalar invariant quantities $m^{2}$, $w^{2}$,
and $s^{2}$, the pseudoscalar quantity $\tilde{s}^{2}$ cannot change
in value under smooth evolution along the system's worldline. To understand
why, observe that if $\tilde{s}^{2}=0$, then it is invariant under
parity and time-reversal transformations, and therefore has the unique
value $\tilde{s}^{2}=0$ for the system's entire phase space. By contrast,
if $\tilde{s}^{2}\ne0$, then our transitive group action of the Poincaré
group can contain only the values $\pm\tilde{s}^{2}$, and no smooth
evolution can take the system from $\tilde{s}^{2}>0$ to $\tilde{s}^{2}<0$
or vice versa. (In all our examples, ahead, we will end up finding
that $\tilde{s}^{2}=0$.)

Classifying the possible systems whose phase spaces provide transitive
group actions of the Poincaré group now reduces to selecting mutually
consistent values for the invariant quantities $m^{2}$, $w^{2}$,
$s^{2}$, and $\tilde{s}^{2}$, and then choosing a convenient reference
state $\parens{X_{\refvalue},p_{\refvalue},S_{\refvalue}}$ that is
compatible with those fixed values. Note again that the constancy
of $m^{2}$, $w^{2}$, $s^{2}$, and $\tilde{s}^{2}$\textemdash including
the constancy of the system's invariant spin-squared $s^{2}$\textemdash is
entirely classical and has nothing to do with quantization or quantum
theory.

As an aside, observe that the only other candidates for invariant
quantities that are derivable from the system's phase-space variables
are 
\begin{align*}
p_{\mu}W^{\mu} & =0,\\
p_{\mu}p_{\nu}S^{\mu\nu} & =0,\\
W_{\mu}W_{\nu}S^{\mu\nu} & =0,\\
W_{\mu}p_{\nu}S^{\mu\nu} & =m^{2}c^{2}\tilde{s}^{2},\\
\epsilon^{\mu\nu\rho\sigma}W_{\mu}p_{\nu}S_{\rho\sigma} & =-2w^{2}.
\end{align*}
 None of these expressions represent fundamentally new quantities
independent of $m^{2}$, $w^{2}$, $s^{2}$, and $\tilde{s}^{2}$,
so we do not need to specify values for them as part of the definition
of our transitive group action of the Poincaré group. 

\subsection{The Generators of the Lorentz Group}

Observe that the system's phase space \eqref{eq:PhaseSpacePointFromPoincTransfFromRef}
is fully parametrized by the values $a^{\mu}$ and $\tud{\Lambda}{\mu}{\nu}$
that make up a generic Poincaré transformation $\parens{a,\Lambda}$,
where $a^{\mu}$ encodes the system's spacetime location and $\tud{\Lambda}{\mu}{\nu}$
encodes the system's motion and angular orientation. Lorentz-transformation
matrices are difficult to manipulate directly, due to the constraint
$\Lambda^{\transp}\eta\Lambda=\eta$ from \eqref{eq:LorentzTransfsPreserveMinkMetric},
so we will find it useful to decompose them into simpler ingredients.\footnote{For a more extensive review of the mathematical details ahead, see
\citep{Weinberg:1996tqtfi}.}

We start by considering a Lorentz transformation $\Lambda\parens{\epsilon}=1-\epsilon$
that differs only infinitesimally from the identity matrix $1$: 
\begin{equation}
\tud{\Lambda}{\alpha}{\beta}\parens{\epsilon}=\delta_{\beta}^{\alpha}-\tud{\epsilon}{\alpha}{\beta}.\label{eq:InfinitesimalLorentzTransf}
\end{equation}
  (The minus sign is conventional.) Here $\tud{\epsilon}{\alpha}{\beta}$
represents a collection of infinitesimal parameters, and $\delta_{\beta}^{\alpha}$
is the four-dimensional Kronecker delta, 
\begin{equation}
\delta_{\beta}^{\alpha}\defeq\begin{cases}
1 & \textrm{for \ensuremath{\alpha=\beta}},\\
0 & \textrm{for \ensuremath{\alpha\ne\beta}},
\end{cases}\label{eq:Def4DKroneckerDelta}
\end{equation}
 which represents the components of the identity matrix. The constraint
$\Lambda^{\transp}\eta\Lambda=\eta$ then yields the equation 
\[
\parens{\delta_{\beta}^{\alpha}-\tud{\epsilon}{\alpha}{\beta}}\eta_{\alpha\gamma}\parens{\delta_{\delta}^{\gamma}-\tud{\epsilon}{\gamma}{\delta}}=\eta_{\beta\delta}.
\]
 Working to first order in $\epsilon$, we see from this equation
that the infinitesimal tensor $\epsilon^{\alpha\beta}$ obtained from
$\tud{\epsilon}{\alpha}{\beta}$ by raising its second index using
the Minkowski metric tensor is antisymmetric: 
\begin{equation}
\epsilon^{\alpha\beta}=-\epsilon^{\beta\alpha}.\label{eq:InfinitesimalLorentzParamMatrixAntisymm}
\end{equation}
 The tensor $\epsilon^{\alpha\beta}$ therefore has six independent
components, with $\epsilon^{yz},\epsilon^{zx},\epsilon^{xy}$ respectively
parametrizing rotations around the $x,y,z$ axes, and with $\epsilon^{tx},\epsilon^{ty},\epsilon^{tz}$
respectively parametrizing Lorentz boosts along the $x,y,z$ directions.

We can write any two-index, antisymmetric Lorentz tensor $A^{\alpha\beta}=-A^{\beta\alpha}$
as 
\begin{align*}
A^{\alpha\beta} & =\frac{1}{2}\parens{A^{\alpha\beta}-A^{\beta\alpha}}\\
 & =\frac{1}{2}A^{\mu\nu}\parens{\delta_{\mu}^{\alpha}\delta_{\nu}^{\beta}-\delta_{\mu}^{\beta}\delta_{\nu}^{\alpha}}.
\end{align*}
 Hence, the tensors defined by 
\begin{equation}
\bracks{\sigma_{\mu\nu}}^{\alpha\beta}\defeq-i\delta_{\mu}^{\alpha}\delta_{\nu}^{\beta}+i\delta_{\mu}^{\beta}\delta_{\nu}^{\alpha}\label{eq:DefLorentzGenerators}
\end{equation}
 form a basis for all two-index, antisymmetric tensors: 
\begin{equation}
A^{\alpha\beta}=\frac{i}{2}A^{\mu\nu}\bracks{\sigma_{\mu\nu}}^{\alpha\beta}.\label{eq:AntisymmTensorFromLorentzGeneratorsBasis}
\end{equation}
 We can therefore write our infinitesimal Lorentz transformation \eqref{eq:InfinitesimalLorentzTransf}
as 
\begin{equation}
\tud{\Lambda}{\alpha}{\beta}\parens{\epsilon}=\delta_{\beta}^{\alpha}-\frac{i}{2}\epsilon^{\mu\nu}\tud{\bracks{\sigma_{\mu\nu}}}{\alpha}{\beta}.\label{eq:InfinitesimalLorentzTransfFromGeneratorsIndices}
\end{equation}
Equivalently, in matrix notation, with the free indices $\alpha$
and $\beta$ suppressed, we can write 
\begin{equation}
\Lambda\parens{\epsilon}=1-\frac{i}{2}\epsilon^{\mu\nu}\sigma_{\mu\nu}.\label{eq:InfinitesimalLorentzTransfFromGenerators}
\end{equation}

The tensors $\tud{\bracks{\sigma_{\mu\nu}}}{\alpha}{\beta}$ are called
the Lorentz generators and are obtained by lowering the $\beta$
index in the definition \eqref{eq:DefLorentzGenerators} using the
Minkowski metric tensor: 
\begin{equation}
\tud{\bracks{\sigma_{\mu\nu}}}{\alpha}{\beta}=-i\delta_{\mu}^{\alpha}\eta_{\nu\beta}+i\eta_{\mu\beta}\delta_{\nu}^{\alpha}.\label{eq:LorentzGeneratorsMixedIndices}
\end{equation}
 We will often suppress the ``additional'' $\alpha,\beta$ indices
for notational economy.

Note that with our overall sign convention for \eqref{eq:LorentzGeneratorsMixedIndices},
the Lorentz generators describe \emph{passive} Lorentz transformations,
which transform our spacetime coordinate axes. If we instead wished
to describe \emph{active} Lorentz transformations, then we could either
replace $\sigma_{\mu\nu}\mapsto-\sigma_{\mu\nu}$ or $\epsilon^{\mu\nu}\mapsto-\epsilon^{\mu\nu}$.

By straightforward calculations, one can show that the Lorentz generators
satisfy the commutation relations 
\begin{align}
 & \bracks{\sigma_{\mu\nu},\sigma_{\rho\sigma}}\defeq\sigma_{\mu\nu}\sigma_{\rho\sigma}-\sigma_{\rho\sigma}\sigma_{\mu\nu}\nonumber \\
 & \qquad=i\eta_{\mu\rho}\sigma_{\nu\sigma}-i\eta_{\mu\sigma}\sigma_{\nu\rho}-i\eta_{\nu\rho}\sigma_{\mu\sigma}+i\eta_{\nu\sigma}\sigma_{\mu\rho},\label{eq:LorentzGeneratorsCommutator}
\end{align}
 and that the matrix product of two Lorentz generators $\sigma_{\mu\nu}$
and $\sigma_{\rho\sigma}$ on their additional $\alpha,\beta$ indices,
traced over those additional indices, yields 
\begin{align}
\frac{1}{2}\Trace\bracks{\sigma^{\mu\nu}\sigma_{\rho\sigma}} & \defeq\frac{1}{2}\tud{\bracks{\sigma^{\mu\nu}}}{\alpha}{\beta}\tud{\bracks{\sigma_{\rho\sigma}}}{\beta}{\alpha}\nonumber \\
 & =\delta_{\rho}^{\mu}\delta_{\sigma}^{\nu}-\delta_{\sigma}^{\mu}\delta_{\rho}^{\nu}\label{eq:LorentzGeneratorsTraceProd}\\
 & =i\bracks{\sigma_{\rho\sigma}}^{\mu\nu}.\label{eq:LorentzGeneratorsTraceProdAlt}
\end{align}
 This last formula implies that antisymmetric tensors $A^{\mu\nu}$
satisfy the identity 
\begin{equation}
\frac{1}{2}\Trace\bracks{\sigma^{\mu\nu}A}=iA^{\mu\nu}.\label{eq:LorentzGeneratorsTracePeelsOffAntisymmTensor}
\end{equation}
Using this formalism, we can rewrite our system's spin tensor \eqref{eq:SpinTensorFromRefIndices}
as 
\begin{align}
S^{\mu\nu} & =-\frac{i}{2}\Trace\bracks{\sigma^{\mu\nu}S}\nonumber \\
 & =-\frac{i}{2}\Trace\bracks{\sigma^{\mu\nu}\Lambda S_{\refvalue}\Lambda^{-1}}.\label{eq:SpinTensorFromTraceLorentzGenerators}
\end{align}

\subsection{The Manifestly Covariant Action Functional}

In the absence of spin, the system's manifestly covariant action functional
takes the form \eqref{eq:ReparamActionFunctionalFromCanonicalMomenta}:
\begin{align}
S_{\textrm{no\,spin}}\bracks{X,\Lambda} & =\int d\lambda\,\mathscr{L}_{\textrm{no\,spin}}\nonumber \\
 & =\int d\lambda\,p_{\mu}\dot{X}^{\mu}=\int d\lambda\,\parens{\Lambda p_{0}}_{\mu}\dot{X}^{\mu}.\label{eq:ActionFunctionalNoSpin}
\end{align}
 Here $X^{\mu}\parens{\lambda}$ and $p^{\mu}\parens{\lambda}\defeq\tud{\Lambda}{\mu}{\nu}\left(\lambda\right)p_{0}^{\nu}$
are functions of the worldline parameter $\lambda$, and dots, as
usual, denote derivatives with respect to $\lambda$. We will eventually
see that this action functional is capable of accommodating particle
types regardless of their mass\textemdash and, in particular, works
just as well for massless particles as it does for particles with
nonzero mass. We will ultimately also need to establish a definite
relationship between the system's four-momentum $p^{\mu}\parens{\lambda}$
and the system's four-velocity $\dot{X}^{\mu}\left(\lambda\right)\defeq dX^{\mu}\left(\lambda\right)/d\lambda$.

In order to include spin in the system's action functional, we will
need to develop a framework for taking derivatives of the variable
Lorentz-transformation matrix $\tud{\Lambda}{\mu}{\nu}\parens{\lambda}$
with respect to the worldline parameter $\lambda$ in a manner that
is consistent with the constraint $\Lambda^{\transp}\eta\Lambda=\eta$
from \eqref{eq:LorentzTransfsPreserveMinkMetric}. To this end, we
examine what happens if we shift slightly forward along the system's
worldline, so that
\begin{equation}
\lambda\to\lambda+d\lambda.\label{eq:InfinitesimalWorldlineParamShift}
\end{equation}
 Using the fact that successive Lorentz transformations compose, 
\begin{equation}
\Lambda^{\prime\prime}=\Lambda^{\prime}\Lambda,\label{eq:LorentzTransformationMatricesCompose}
\end{equation}
 and recalling the formula \eqref{eq:InfinitesimalLorentzTransfFromGenerators}
for a Lorentz transformation that differs infinitesimally from the
identity, with $d\theta^{\mu\nu}\defeq\epsilon^{\mu\nu}$ denoting
our Lorentz-boost and angular parameters, it follows that 
\begin{align}
\Lambda\parens{\lambda+d\lambda} & =\Lambda\parens{d\lambda}\Lambda\parens{\lambda}\nonumber \\
 & =\parens{1-\parens{i/2}d\theta^{\mu\nu}\parens{\lambda}\sigma_{\mu\nu}}\Lambda\parens{\lambda}.\label{eq:InfinitesimalShiftLorentzTransf}
\end{align}
 We can rearrange this formula to obtain the derivative of $\Lambda\parens{\lambda}$
with respect to $\lambda$ in terms of the rates of change $\dot{\theta}^{\mu\nu}\parens{\lambda}\defeq d\theta^{\mu\nu}\parens{\lambda}/d\lambda$:
\begin{align}
\dot{\Lambda}\parens{\lambda} & \defeq\lim_{d\lambda\to0}\frac{\Lambda\parens{\lambda+d\lambda}-\Lambda\parens{\lambda}}{d\lambda}\nonumber \\
 & =-\frac{i}{2}\dot{\theta}^{\mu\nu}\parens{\lambda}\sigma_{\mu\nu}\Lambda\parens{\lambda}.\label{eq:ParamDerivLorentzTransf}
\end{align}
 Hence, 
\[
\dot{\Lambda}\parens{\lambda}\Lambda^{-1}\parens{\lambda}=-\frac{i}{2}\dot{\theta}^{\mu\nu}\parens{\lambda}\sigma_{\mu\nu},
\]
 and so, invoking the identities \eqref{eq:AntisymmTensorFromLorentzGeneratorsBasis}
and \eqref{eq:LorentzGeneratorsTraceProdAlt}, we obtain an important
formula for the rates of change $\dot{\theta}^{\mu\nu}\parens{\lambda}$
of the Lorentz-transformation parameters: 
\begin{equation}
\dot{\theta}^{\mu\nu}\parens{\lambda}=\frac{i}{2}\Trace\bracks{\sigma^{\mu\nu}\dot{\Lambda}\parens{\lambda}\Lambda^{-1}\parens{\lambda}}.\label{eq:ParamDerivFromTrace}
\end{equation}
 Despite the factor of $i$, this expression is purely real, due to
the additional factor of $i$ in the definition \eqref{eq:DefLorentzGenerators}
of $\sigma^{\mu\nu}$. 

We now look back at the manifestly covariant Lagrangian appearing
as the integrand of our action functional \eqref{eq:ActionFunctionalNoSpin}:
\begin{equation}
\mathscr{L}_{\textrm{no\,spin}}=p_{\mu}\dot{X}^{\mu}.\label{eq:CovariantLagrangianNoSpin}
\end{equation}
 Using the product rule in reverse (that is, ``integration by parts''
without an actual integration), we can move the derivative from $X^{\mu}\parens{\lambda}$
to $p_{\mu}\parens{\lambda}$ at the cost of an overall minus sign
and an additive total derivative that does not affect the system's
equations of motion. The result is 
\[
\mathscr{L}_{\textrm{no\,spin}}=-X_{\mu}\dot{p}^{\mu}+\parens{\textrm{total derivative}}.
\]
 Remembering that the system's four-momentum $p^{\mu}\parens{\lambda}$
here is fundamentally defined according to \eqref{eq:FourMomentumFromRefIndices}
in terms of its fixed reference value $p_{\refvalue}^{\mu}$ and the
variable Lorentz-transformation matrix $\tud{\Lambda}{\mu}{\nu}\parens{\lambda}$,
\[
p^{\mu}\parens{\lambda}\defeq\tud{\Lambda}{\mu}{\nu}\parens{\lambda}p_{\refvalue}^{\nu},
\]
 and relabeling indices for later convenience, we have 
\[
\mathscr{L}_{\textrm{no\,spin}}=-X_{\alpha}\tud{\dot{\Lambda}}{\alpha}{\gamma}p_{\refvalue}^{\gamma}+\parens{\textrm{total derivative}}.
\]
 Invoking \eqref{eq:ParamDerivLorentzTransf} for the derivative of
the Lorentz-transformation matrix yields 
\begin{align*}
\mathscr{L}_{\textrm{no\,spin}} & =-X_{\alpha}\biggparens{-\frac{i}{2}\dot{\theta}^{\mu\nu}\tud{\bracks{\sigma_{\mu\nu}}}{\alpha}{\beta}\tud{\Lambda}{\beta}{\gamma}}p_{\refvalue}^{\gamma}\\
 & \qquad\qquad\qquad+\parens{\textrm{total derivative}}\\
 & =\frac{1}{2}X_{\alpha}i\tud{\bracks{\sigma_{\mu\nu}}}{\alpha}{\beta}p^{\beta}\dot{\theta}^{\mu\nu}+\parens{\textrm{total derivative}}.
\end{align*}
Recalling our formula \eqref{eq:LorentzGeneratorsMixedIndices} for
the Lorentz generators $\tud{\bracks{\sigma_{\mu\nu}}}{\alpha}{\beta}$,
this expression simplifies to 
\begin{align*}
\mathscr{L}_{\textrm{no\,spin}} & =\frac{1}{2}X_{\alpha}\parens{\delta_{\mu}^{\alpha}\eta_{\nu\beta}-\eta_{\mu\beta}\delta_{\nu}^{\alpha}}p^{\beta}\dot{\theta}^{\mu\nu}\\
 & \qquad\qquad\qquad+\parens{\textrm{total derivative}}\\
 & =\frac{1}{2}\parens{X_{\mu}p_{\nu}-X_{\nu}p_{\mu}}\dot{\theta}^{\mu\nu}+\parens{\textrm{total derivative}}.
\end{align*}
The quantity in parentheses is precisely the system's orbital angular-momentum
tensor $L_{\mu\nu}$, as defined in \eqref{eq:Def4DOrbitalAngularMomentum},
so we end up with 
\begin{equation}
\mathscr{L}_{\textrm{no\,spin}}=\frac{1}{2}L_{\mu\nu}\dot{\theta}^{\mu\nu}+\parens{\textrm{total derivative}}.\label{eq:CovariantLagrangianNoSpinFromOrbAngMom}
\end{equation}

The first term in \eqref{eq:CovariantLagrangianNoSpinFromOrbAngMom}
has precisely the form of a canonical momentum contracted with the
rates of change of its corresponding canonical coordinates, where
the factor of $1/2$ naturally prevents the implicit summation from
double-counting independent terms in the contraction of the two antisymmetric
tensors $L_{\mu\nu}=-L_{\nu\mu}$ and $\dot{\theta}^{\mu\nu}=-\dot{\theta}^{\nu\mu}$.
It may seem surprising that we have managed to rewrite the system's
kinetic Lagrangian $\mathscr{L}_{\textrm{no\,spin}}=p_{\mu}\dot{X}^{\mu}$
in terms of what looks superficially like purely orbital angular momentum,
but remember that the temporal components $L^{ti}$ of the orbital
angular-momentum tensor are not angular momenta\textemdash indeed,
in light of \eqref{eq:OrbAngMomTemporalCompsAsVelFormula}, they actually
encode linear motion.

Including the system's spin in the dynamics means generalizing the
orbital angular-momentum tensor $L_{\mu\nu}$ in \eqref{eq:CovariantLagrangianNoSpinFromOrbAngMom}
to the \emph{total} angular-momentum tensor $J_{\mu\nu}$ defined
in \eqref{eq:Def4DTotalAngMomTensor}, 
\[
L_{\mu\nu}\mapsto J_{\mu\nu}\defeq L_{\mu\nu}+S_{\mu\nu},
\]
 where $S_{\mu\nu}$ is the system's spin tensor. The system's manifestly
covariant Lagrangian correspondingly becomes 
\begin{align}
 & \mathscr{L}_{\textrm{no\,spin}}\mapsto\mathscr{L}\defeq\frac{1}{2}J_{\mu\nu}\dot{\theta}^{\mu\nu}+\parens{\textrm{total derivative}}\nonumber \\
 & \qquad\qquad=\frac{1}{2}L_{\mu\nu}\dot{\theta}^{\mu\nu}+\frac{1}{2}S_{\mu\nu}\dot{\theta}^{\mu\nu}+\parens{\textrm{total derivative}}.\label{eq:CovariantLagrangianFromTotAngMom}
\end{align}

At this point, we are free to recombine the first and last terms in
$\mathscr{L}$ to get back the expression $p_{\mu}\dot{X}^{\mu}$
that we started with. Moreover, by contracting both sides of our formula
\eqref{eq:ParamDerivFromTrace} for $\dot{\theta}^{\mu\nu}$ with
the system's spin tensor $S^{\mu\nu}$, and using $\parens{i/2}S_{\mu\nu}\tud{\bracks{\sigma^{\mu\nu}}}{\alpha}{\beta}=\tud S{\alpha}{\beta}$
from \eqref{eq:AntisymmTensorFromLorentzGeneratorsBasis}, we can
write the second term in \eqref{eq:CovariantLagrangianFromTotAngMom}
as 
\begin{equation}
\frac{1}{2}S_{\mu\nu}\parens{\lambda}\dot{\theta}^{\mu\nu}\parens{\lambda}=\frac{1}{2}\Trace\bracks{S\parens{\lambda}\dot{\Lambda}\parens{\lambda}\Lambda^{-1}\parens{\lambda}}.\label{eq:SpinKineticTerm}
\end{equation}
  Hence, as originally shown in \citep{HansonRegge:1974rst,SkagerstamStern:1981ldccps,BalachandranMarmoSkagerstamStern:1983gsfb,Frydryszak:1996lmpsfsy},
the complete action functional for the system is 
\begin{equation}
S\bracks{X,\Lambda}=\int d\lambda\,\mathscr{L}=\int d\lambda\,\biggparens{p_{\mu}\dot{X}^{\mu}+\frac{1}{2}\Trace\bracks{S\dot{\Lambda}\Lambda^{-1}}}.\label{eq:ActionFunctionalWithSpin}
\end{equation}

In using the action functional \eqref{eq:ActionFunctionalWithSpin},
keep in mind that the four-momentum $p^{\mu}\parens{\lambda}$ and
the spin tensor $S^{\mu\nu}\parens{\lambda}$ are given respectively
by \eqref{eq:FourMomentumFromRefIndices} and \eqref{eq:SpinTensorFromTraceLorentzGenerators}
in terms of their constant reference values $p_{\refvalue}^{\mu}$
and $S_{\refvalue}^{\mu\nu}$ together with the variable Lorentz-transformation
matrix $\tud{\Lambda}{\mu}{\nu}\parens{\lambda}$: 
\begin{align}
p^{\mu}\parens{\lambda} & \defeq\tud{\Lambda}{\mu}{\nu}\parens{\lambda}p_{\refvalue}^{\nu},\label{eq:4MomFromRefParam}\\
S^{\mu\nu}\parens{\lambda} & \defeq\tud{\Lambda}{\mu}{\rho}\parens{\lambda}S_{\refvalue}^{\rho\sigma}\tdu{\parens{\Lambda^{\transp}}}{\sigma}{\nu}\parens{\lambda}\nonumber \\
 & =-\frac{i}{2}\Trace\bracks{\sigma^{\mu\nu}\Lambda\parens{\lambda}S_{\refvalue}\Lambda^{-1}\parens{\lambda}}.\label{eq:SpinTensorFromRefParam}
\end{align}
 Consequently, before the equations of motion are imposed, neither
$p^{\mu}\parens{\lambda}$ nor $S^{\mu\nu}\parens{\lambda}$ depend
on the spacetime degrees of freedom $X^{\mu}\parens{\lambda}$.

\subsection{The Equations of Motion}

To obtain the system's equations of motion, we apply the extremization
condition \eqref{eq:VariationActionFunctional} by varying the action
functional \eqref{eq:ActionFunctionalWithSpin} with respect to its
fundamental variables $X^{\mu}$ and $\tud{\Lambda}{\mu}{\nu}$. The
spin term $\parens{1/2}\Trace\bracks{S\dot{\Lambda}\Lambda^{-1}}$
does not involve the spacetime coordinates $X^{\mu}$, so varying
the action functional with respect to $X^{\mu}$ yields 
\begin{align*}
\delta_{X}S & =\int d\lambda\,\parens{p_{\mu}\delta\dot{X}^{\mu}+0}\\
 & =\int d\lambda\,p_{\mu}\frac{d}{d\lambda}\delta X^{\mu}\\
 & =-\int d\lambda\,\dot{p}_{\mu}\delta X^{\mu},
\end{align*}
 where we have dropped a boundary term. Setting this variation equal
to zero for arbitrary $\delta X^{\mu}$ leads to the system's first
equation of motion, which we see describes conservation of energy-momentum:
\begin{equation}
\dot{p}^{\mu}=0.\label{eq:4MomEOM}
\end{equation}
 Notice that this equation of motion, by itself, does not determine
the system's four-velocity $\dot{X}^{\mu}\defeq dX^{\mu}/d\lambda$,
or even establish any sort of relationship between $p^{\mu}$ and
$\dot{X}^{\mu}$. We will return to this issue later.

Varying the action functional with respect to the variable Lorentz-transformation
matrix $\tud{\Lambda}{\mu}{\nu}$ is more complicated, due to its
appearance in both terms in the integrand. As our first step, we
find 
\begin{equation}
\delta_{\Lambda}S=\int d\lambda\,\biggparens{\parens{\delta p^{\mu}}\dot{X}_{\mu}+\frac{1}{2}\Trace\bracks{\delta\parens{S\dot{\Lambda}\Lambda^{-1}}}}.\label{eq:VaryActionFunctionalLorentzMatrixIntermed}
\end{equation}
 Invoking our formula \eqref{eq:4MomFromRefParam} for the four-momentum
$p^{\mu}$ in terms of its reference value $p_{\refvalue}^{\mu}$
and the Lorentz-transformation matrix $\tud{\Lambda}{\mu}{\nu}$,
the first term in \eqref{eq:VaryActionFunctionalLorentzMatrixIntermed}
gives 
\begin{align*}
\parens{\delta p^{\mu}}\dot{X}_{\mu} & =\parens{\delta\tud{\Lambda}{\mu}{\nu}}p_{\refvalue}^{\nu}\dot{X}_{\mu}\\
 & =\tud{\parens{-\parens{i/2}\delta\theta^{\rho\sigma}\sigma_{\rho\sigma}\Lambda}}{\mu}{\nu}p_{\refvalue}^{\nu}\dot{X}_{\mu}\\
 & =-\frac{i}{2}\delta\theta^{\rho\sigma}\tud{\bracks{\sigma_{\rho\sigma}}}{\mu}{\nu}p^{\nu}\dot{X}_{\mu}\\
 & =-\frac{i}{2}\delta\theta^{\rho\sigma}\parens{-i\delta_{\rho}^{\mu}\eta_{\sigma\nu}+i\eta_{\rho\nu}\delta_{\sigma}^{\mu}}p^{\nu}\dot{X}_{\mu}\\
 & =\frac{1}{2}\parens{-\dot{X}_{\rho}p_{\sigma}+\dot{X}_{\sigma}p_{\rho}}\delta\theta^{\rho\sigma}.
\end{align*}
 Meanwhile, using $\tud S{\alpha}{\beta}=\tud{\parens{\Lambda S_{\refvalue}\Lambda^{-1}}}{\alpha}{\beta}$,
the second term in \eqref{eq:VaryActionFunctionalLorentzMatrixIntermed}
gives 
\begin{align*}
\frac{1}{2}\Trace\bracks{\delta\parens{S\dot{\Lambda}\Lambda^{-1}}} & =\frac{1}{2}\Trace\bracks{S_{\refvalue}\delta\parens{\Lambda^{-1}\dot{\Lambda}}}\\
 & =\frac{1}{2}\Trace\bracks{S_{\refvalue}\delta\parens{\Lambda^{-1}}\dot{\Lambda}+S_{\refvalue}\Lambda^{-1}\delta\dot{\Lambda}}\\
 & =\frac{1}{2}\Trace\bracks{S_{\refvalue}\Lambda^{-1}\parens{-\parens{i/2}\delta\dot{\theta}^{\rho\sigma}\sigma_{\rho\sigma}}\Lambda}\\
 & =-\frac{i}{4}\Trace\bracks{S_{\refvalue}\Lambda^{-1}\sigma_{\rho\sigma}\Lambda}\delta\dot{\theta}^{\rho\sigma}\\
 & =\frac{1}{2}S_{\rho\sigma}\frac{d}{d\lambda}\delta\theta^{\rho\sigma},
\end{align*}
 where we have invoked \eqref{eq:SpinTensorFromRefParam} in the last
step. Thus, dropping a boundary term, we see that the overall variation
\eqref{eq:VaryActionFunctionalLorentzMatrixIntermed} in the action
functional reduces to 
\[
\delta_{\Lambda}S=\int d\lambda\,\frac{1}{2}\parens{-\dot{X}_{\rho}p_{\sigma}+\dot{X}_{\sigma}p_{\rho}-\dot{S}_{\rho\sigma}}\delta\theta^{\rho\sigma}.
\]
 Setting this variation equal to zero for arbitrary $\delta\theta^{\rho\sigma}$
leads to the system's second equation of motion: 
\begin{equation}
\dot{S}^{\mu\nu}=-\dot{X}^{\mu}p^{\nu}+\dot{X}^{\nu}p^{\mu}.\label{eq:SpinTensorEOM}
\end{equation}

To provide an interpretation for this equation of motion, we recall
again the definition \eqref{eq:Def4DOrbitalAngularMomentum} of the
tensor $L^{\mu\nu}$ that encodes the system's orbital angular momentum:
\[
L^{\mu\nu}\defeq X^{\mu}p^{\nu}-X^{\nu}p^{\mu}.
\]
 Because the system's four-momentum $p^{\mu}$ is conserved, \eqref{eq:4MomEOM},
we see that the rate of change in $L^{\mu\nu}$ is given by 
\begin{equation}
\dot{L}^{\mu\nu}=\dot{X}^{\mu}p^{\nu}-\dot{X}^{\nu}p^{\mu},\label{eq:RateOfChangeOrbAngMom}
\end{equation}
 so we can recast the equation of motion \eqref{eq:SpinTensorEOM}
for the spin tensor $S^{\mu\nu}$ as the statement that the system's
total angular momentum $J^{\mu\nu}\defeq L^{\mu\nu}+S^{\mu\nu}$ is
conserved: 
\begin{equation}
\dot{J}^{\mu\nu}=0.\label{eq:ConservationTotalAngMom}
\end{equation}
 Combining $\dot{p}^{\mu}=0$ and $\dot{J}^{\mu}=0$, it follows
immediately that the system's Pauli-Lubanski pseudovector \eqref{eq:DefPauliLubanski4Vec}
is likewise constant in time: 
\begin{equation}
\dot{W}^{\mu}=0.\label{eq:PauliLubanskiConstant}
\end{equation}

At a deeper level, the system's two equations of motion \eqref{eq:4MomEOM},
$\dot{p}^{\mu}=0$, and \eqref{eq:ConservationTotalAngMom}, $\dot{J}^{\mu\nu}=0$,
are consequences of Noether's theorem together with the fact that
the system's action functional \eqref{eq:ActionFunctionalWithSpin}
has continuous symmetries under spacetime translations and Lorentz
transformations.

\subsection{Self-Consistency Conditions on the Phase Space}

Now that we know the system's equations of motion, we will need to
ensure that they are consistent with the invariance of the fixed quantities
$m^{2}$, $w^{2}$, $s^{2}$, and $\tilde{s}^{2}$ from \eqref{eq:Def4DMassSquaredAsInvariant}\textendash \eqref{eq:Def4DDualSpinSquaredInvariant}.

For our first check of self-consistency, we note that the invariance
of $p^{2}\defeq-m^{2}c^{2}$ is compatible with the equation of motion
\eqref{eq:4MomEOM}, $\dot{p}^{\mu}=0$: 
\begin{equation}
\frac{d}{d\lambda}\parens{p^{2}}=2p_{\mu}\dot{p}^{\mu}=0.\label{eq:CheckMassSqConst}
\end{equation}
 Similarly, the constancy of $W^{2}\defeq w^{2}$ is compatible with
the constancy \eqref{eq:PauliLubanskiConstant} of the Pauli-Lubanski
pseudovector: 
\begin{equation}
\frac{d}{d\lambda}\parens{W^{2}}=2W_{\mu}\dot{W}^{\mu}=0.\label{eq:CheckPauliLubanskiSquaredConst}
\end{equation}

By contrast, the constancy of the spin-squared scalar $\parens{1/2}S_{\mu\nu}S^{\mu\nu}\defeq s^{2}$,
combined with the equation of motion \eqref{eq:SpinTensorEOM}, $\dot{S}^{\mu\nu}=-\dot{X}^{\mu}p^{\nu}+\dot{X}^{\nu}p^{\mu}$,
requires that 
\begin{equation}
\frac{d}{d\lambda}\biggparens{\frac{1}{2}S_{\mu\nu}S^{\mu\nu}}=S_{\mu\nu}\dot{S}^{\mu\nu}=2\dot{X}^{\nu}p^{\mu}S_{\mu\nu}=0.\label{eq:DerivSquaredSpinTensorVanishes}
\end{equation}
 Again, keep in mind that we have not yet established a definite relationship
between the system's four-momentum $p^{\mu}$ and its four-velocity
$\dot{X}^{\mu}\defeq dX^{\mu}/d\lambda$. In particular, it is not
clear at this point whether or not $p^{\mu}$ is proportional to $\dot{X}^{\mu}$,
so the condition \eqref{eq:DerivSquaredSpinTensorVanishes} is not
trivial.

Because the condition \eqref{eq:DerivSquaredSpinTensorVanishes} must
hold for all solution trajectories, it imposes an additional requirement
on the system's phase space. Specifically, the system's reference
four-momentum $p_{\refvalue}^{\mu}$ and its reference spin tensor
$S_{\refvalue}^{\mu\nu}$ must satisfy 
\begin{equation}
p_{\refvalue,\mu}S_{\refvalue}^{\mu\nu}=0.\label{eq:RefFourMomSpinTensorZero}
\end{equation}
 The tensor-contraction appearing on the left-hand side therefore
vanishes in one inertial reference frame, so it remains zero under
all Poincaré transformations and therefore represents a Poincaré-invariant
statement about the system's phase space: 
\begin{equation}
p_{\mu}S^{\mu\nu}=0.\label{eq:FourMomSpinTensorZeroPhysicalCondition}
\end{equation}

Notice that this last self-consistency condition, which was also introduced
in \citep{SkagerstamStern:1981ldccps}, is phrased in terms of linear
and angular momentum, both of which have physical meanings in classical
field theory. In particular, the condition $p_{\mu}S^{\mu\nu}=0$
is closely related to the momentum-space version of the Lorenz equation
\begin{equation}
\partial_{\mu}A^{\mu}=0\label{eq:LorenzEquation}
\end{equation}
 that appears both in the Proca theory of a massive spin-1 bosonic
field and as the condition for Lorenz gauge in electromagnetism. Like
the Lorenz equation in those field theories, we will eventually see
that the condition \eqref{eq:FourMomSpinTensorZeroPhysicalCondition}
ends up eliminating unphysical spin states.

To make this connection with field theory more explicit, consider
the classical theory of a spin-1 field, and let 
\begin{equation}
A^{\mu}\left(x\right)=\varepsilon^{\mu}\exp\left(ip\dotprod x/\hbar\right)\label{eq:Spin1FieldPlaneWave}
\end{equation}
 be a monochromatic plane wave of the field. Here $\varepsilon^{\mu}$
is the wave's polarization four-vector, which ultimately encodes the
field's quantized spin. Then the Lorenz equation $\partial_{\mu}A^{\mu}=0$
reduces to the statement that 
\begin{equation}
p_{\mu}\varepsilon^{\mu}=0,\label{eq:MomentumLorenzEquation}
\end{equation}
 which eliminates one linear combination of polarizations, meaning
one unphysical spin state. In contrast with a spin-1 field theory,
our classical particle's invariant spin is not quantized, and our
condition $p_{\mu}S^{\mu\nu}=0$ will turn out to eliminate a continuous
infinity of unphysical spin states.

Our final self-consistency condition is that the derivative of the
pseudoscalar invariant quantity $\parens{1/8}\epsilon_{\mu\nu\rho\sigma}S^{\mu\nu}S^{\rho\sigma}\defeq\tilde{s}^{2}$
must vanish: 
\begin{align}
\frac{d}{d\lambda}\biggparens{\frac{1}{8}\epsilon_{\mu\nu\rho\sigma}S^{\mu\nu}S^{\rho\sigma}} & =\frac{1}{4}\epsilon_{\mu\nu\rho\sigma}\dot{S}^{\mu\nu}S^{\rho\sigma}\nonumber \\
 & =-\frac{1}{2}\epsilon_{\mu\nu\rho\sigma}\dot{X}^{\mu}p^{\nu}S^{\rho\sigma}\nonumber \\
 & =\dot{X}^{\mu}W_{\mu}=0.\label{eq:DerivDualSquaredSpinTensorVanishesCondition}
\end{align}
 We will need to verify in the explicit examples ahead that this condition
is indeed satisfied. 

\subsection{The Four-Velocity}

The self-consistency condition \eqref{eq:FourMomSpinTensorZeroPhysicalCondition},
$p_{\mu}S^{\mu\nu}=0$, will play an important role in our work ahead.
As we will now investigate, its implications include a general set
of relationships between the system's four-momentum $p^{\mu}$ and
its four-velocity $\dot{X}^{\mu}$.

Taking a derivative of both sides of $p_{\mu}S^{\mu\nu}=0$ with respect
to the worldline parameter $\lambda$, and invoking the equations
of motion \eqref{eq:4MomEOM}, $\dot{p}^{\mu}=0$, and \eqref{eq:SpinTensorEOM},
$\dot{S}^{\mu\nu}=-\dot{X}^{\mu}p^{\nu}+\dot{X}^{\nu}p^{\mu}$, we
obtain 
\begin{equation}
p_{\mu}\dot{S}^{\mu\nu}=-\parens{p\dotprod\dot{X}}p^{\nu}+\parens{-m^{2}c^{2}}\dot{X}^{\nu}=0,\label{eq:ContractionFourMomDerivSpinTensor}
\end{equation}
 which gives us an equation that relates $p^{\mu}$ and $\dot{X}^{\mu}$:
\begin{equation}
\parens{p\dotprod\dot{X}}p^{\mu}=\parens{-m^{2}c^{2}}\dot{X}^{\mu}.\label{eq:FourMomFourVelRelationship}
\end{equation}
 Contracting both sides with $\dot{X}_{\mu}$, we find 
\begin{equation}
\parens{p\dotprod\dot{X}}^{2}=m^{2}c^{2}\parens{-\dot{X}^{2}}.\label{eq:FourMomFourVelSquaredRelation}
\end{equation}
 Taking the square root of this last equation, and substituting the
result back into the equation before it, we arrive at the following
pair of equations: 
\begin{align}
p\dotprod\dot{X} & =\pm mc^{2}\sqrt{-\dot{X}^{2}/c^{2}},\label{eq:FourMomDotProdFourVelRelationship}\\
m\sqrt{-\dot{X}^{2}/c^{2}}\,p^{\mu} & =\mp m^{2}\dot{X}^{\mu}.\label{eq:FourMomFourVelSqrRootRelationship}
\end{align}
 Together with \eqref{eq:4MomEOM}, $\dot{p}^{\mu}=0$, \eqref{eq:ConservationTotalAngMom},
$\dot{J}^{\mu\nu}=0$, and \eqref{eq:FourMomSpinTensorZeroPhysicalCondition},
$p_{\mu}S^{\mu\nu}=0$, these final two equations complete our specification
of the system's dynamics.

\section{Classification of the Transitive Group Actions of the Orthochronous
Poincaré Group}

We are now ready to apply the foregoing framework to the task of classifying
systems whose phase spaces provide transitive group actions of the
Poincaré group. For simplicity, we will focus our attention on transitive
group actions of the \emph{orthochronous} Poincaré group, putting
aside time-reversal transformations \eqref{eq:TimeReversalTransf}
until our paper's conclusion.

Notice then that for $m^{2}\geq0$, the system's four-momentum $p^{\mu}$
is either timelike or null, $p^{2}\leq0$, and so \eqref{eq:CausalPropertySignTemporalComponentTimelikeNull4Vecs}
implies that the \emph{sign} of $p^{\mu}$ is an invariant property
of the system. When we consider transitive group actions having $m^{2}\geq0$,
we will assume the positive-energy case $p^{t}>0$ on physical grounds.
We will address the ``negative-energy'' case $p^{t}<0$ in our conclusion.

\subsection{Massive, Positive-Energy Particles}

As our first example, we consider a transitive group action of the
orthochronous Poincaré group for which $m>0$ is real and positive,
and for which the system's energy $E=p^{t}c>0$ is likewise positive.
Then $p^{\mu}$ is a timelike four-vector, so we know from \eqref{eq:CausalPropertySignTemporalComponentTimelikeNull4Vecs}
that the sign of $p^{t}$ is invariant under orthochronous Lorentz
transformations, and thus our choice of positive energy is well-defined.

Given that $p^{2}=-m^{2}c^{2}$ for $m>0$ with positive $p^{t}$,
we can express the system's energy $E=p^{t}c$ in terms of its three-dimensional
momentum $\vec p=\parens{p_{x},p_{y},p_{z}}$ as 
\begin{equation}
E=\sqrt{\vec p^{2}c^{2}+m^{2}c^{4}},\label{eq:MassivePosEnergyMassShellRelation}
\end{equation}
 a formula known as the system's mass-shell relation because it takes
the visual form of a hyperboloid (a ``shell'') when plotted in terms
of the four variables $E$, $p_{x}$, $p_{y}$, $p_{z}$. Furthermore,
there exists a state of the system in which the four-momentum $p^{\mu}$
takes the specific value $\parens{mc,\vec 0}^{\mu}$, which we will
choose to be its reference value: 
\begin{equation}
p_{\refvalue}^{\mu}\defeq\parens{mc,\vec 0}^{\mu}=mc\,\delta_{t}^{\mu}.\label{eq:DefMassivePosEnergyRef4Mom}
\end{equation}

Due to the condition $m>0$, the four-momentum $p^{\mu}$ cannot vanish,
and under our assumption of a strictly monotonic parametrization $X^{\mu}\parens{\lambda}$,
the four-velocity $\dot{X}^{\mu}$ cannot vanish either, so the relation
\eqref{eq:FourMomFourVelSqrRootRelationship}, 
\[
m\sqrt{-\dot{X}^{2}/c^{2}}\,p^{\mu}=\mp m^{2}\dot{X}^{\mu},
\]
 implies that $\dot{X}^{2}\ne0$. We therefore have 
\[
p^{\mu}=m\frac{\dot{X}^{\mu}}{\sqrt{-\dot{X}^{2}/c^{2}}},
\]
 where we have taken the positive sign by choosing our parametrization
$X^{\mu}\parens{\lambda}$ such that $\dot{X}^{\mu}$ is future-directed.
We therefore learn that the system's four-momentum $p^{\mu}$ is given
by 
\begin{equation}
p^{\mu}=mu^{\mu},\label{eq:MassivePosEnergy4MomFrom4Vel}
\end{equation}
 where $u^{\mu}$ is the system's \emph{normalized} four-velocity:
\begin{equation}
u^{\mu}\defeq\frac{\dot{X}^{\mu}}{\sqrt{-\dot{X}^{2}/c^{2}}},\quad u^{2}=-c^{2}.\label{eq:DefNormalizedFourVel}
\end{equation}

We can interpret the equation \eqref{eq:MassivePosEnergy4MomFrom4Vel}
as supplying our \emph{definition} of $\dot{X}^{\mu}$ (or $u^{\mu}$)
in terms of $p^{\mu}$ and $m$. Furthermore, because $p^{\mu}$ is
parallel to $u^{\mu}$, we see that the self-consistency condition
\eqref{eq:DerivDualSquaredSpinTensorVanishesCondition}, $\dot{X}^{\mu}W_{\mu}=0$,
is satisfied.

As a consequence of \eqref{eq:DefNormalizedFourVel}, we also see
that when the system is in its reference state with $p^{\mu}=p_{\refvalue}^{\mu}=\parens{mc,\vec 0}^{\mu}$,
the four-velocity describes the system at rest, with 
\begin{equation}
u_{\refvalue}^{\mu}=\parens{c,\vec 0}^{\mu}=u_{\textrm{rest}}^{\mu}.\label{eq:MassivePosEnergyRest4Vel}
\end{equation}
 For general states, the equation of motion \eqref{eq:4MomEOM} for
the system's four-momentum, $\dot{p}^{\mu}=0$, tells us that the
system's normalized four-velocity is constant: 
\begin{equation}
\dot{u}^{\mu}=0.\label{eq:MassivePosEnergy4VelConst}
\end{equation}
 It follows that the system describes a pointlike particle that travels
along a straight, timelike path in spacetime.

Defining the particle's three-dimensional velocity $\vec v=\parens{v_{x},v_{y},v_{z}}$
as 
\begin{equation}
\vec v\defeq\frac{d\vec X}{dt}=\frac{\dot{\vec X}}{\dot{T}},\label{eq:Def3DVelFromParamDerivs}
\end{equation}
 and using \eqref{eq:MassivePosEnergy4MomFrom4Vel}, $p^{\mu}=mu^{\mu}$,
together with $E=p^{t}c$ and the mass-shell relation \eqref{eq:MassivePosEnergyMassShellRelation}
between $E$ and $\vec p$, we also obtain an important equation connecting
the system's three-dimensional velocity $\vec v$ and its three-dimensional
momentum $\vec p$: 
\begin{equation}
\vec v=\frac{\vec pc^{2}}{E}=\frac{\vec p}{\verts{\vec p}}\frac{c}{\sqrt{1+m^{2}c^{2}/\vec p^{2}}}.\label{eq:MassivePosEnergy3DMomVelRelation}
\end{equation}
 We see right away from this equation that the particle's speed $\verts{\vec v}$
is always slower than the speed of light $c$: 
\begin{equation}
\verts{\vec v}<c.\label{eq:MassivePosEnergySpeedLessThanSpeedOfLight}
\end{equation}
 Moreover, in the general case in which particle may be in motion,
its normalized four-velocity is 
\begin{equation}
u^{\mu}=\parens{\gamma c,\gamma\vec v}^{\mu},\label{eq:MassivePosEnergyNorm4VelFromGamma}
\end{equation}
 where the Lorentz factor $\gamma$ is defined by 
\begin{equation}
\gamma\defeq\frac{1}{\sqrt{1-\vec v^{2}/c^{2}}}\geq1.\label{eq:DefLorentzFactor}
\end{equation}

We next examine the particle's orbital and spin angular momentum.
The relation \eqref{eq:MassivePosEnergy4MomFrom4Vel}, $p^{\mu}=mu^{\mu}=m\dot{X}^{\mu}/\sqrt{-\dot{X}^{2}/c^{2}}$,
immediately implies that the particle's orbital angular momentum \eqref{eq:Def4DOrbitalAngularMomentum}
is conserved: 
\begin{equation}
\dot{L}^{\mu\nu}=\dot{X}^{\mu}p^{\nu}-\dot{X}^{\nu}p^{\mu}=0.\label{eq:MassivePosEnergyOrbAngMomConst}
\end{equation}
 Remembering our formula \eqref{eq:OrbAngMomTemporalCompsAsVelFormula}
for the temporal components $L^{ti}$ of the orbital angular-momentum
tensor, 
\[
L^{ti}=-\frac{E}{c}\biggparens{X^{i}-\frac{p^{i}c^{2}}{E}T},
\]
 and invoking the constancy of $E$ and $p^{i}$ from the equation
of motion \eqref{eq:4MomEOM} for $p^{\mu}$, we see that $\dot{L}^{ti}=0$
gives the relation 
\[
\frac{\vec pc^{2}}{E}=\frac{\dot{\vec X}}{\dot{T}},
\]
 which is just our earlier equation \eqref{eq:MassivePosEnergy3DMomVelRelation}
connecting the particle's three-dimensional velocity $\vec v$ to
its three-dimensional momentum $\vec p$.\footnote{More generally, for a system of multiple particles labeled by $\alpha=1,2,\dotsc$,
the spatial components $L^{ti}$ generalize to the system's center-of-mass-energy
$X_{\textrm{CM}}^{i}=\sum_{\alpha}E_{\alpha}X_{\textrm{initial},\alpha}^{i}/E_{\textrm{total}}$,
and so their conservation implies the constancy of $\vec X_{\textrm{CM}}$.}

Combining the conservation equation \eqref{eq:MassivePosEnergyOrbAngMomConst}
for the particle's orbital angular-momentum tensor $L^{\mu\nu}$ with
the equation of motion \eqref{eq:SpinTensorEOM} for the particle's
spin tensor $S^{\mu\nu}$ tells us that the particle's spin is separately
conserved: 
\begin{equation}
\dot{S}^{\mu\nu}=0.\label{eq:MassivePosEnergySpinConst}
\end{equation}
 Furthermore, the condition \eqref{eq:RefFourMomSpinTensorZero},
$p_{\refvalue,\mu}S_{\refvalue}^{\mu\nu}=0$, becomes 
\begin{equation}
mc\,S_{\refvalue}^{t\nu}=0,\label{eq:MassivePosEnergyRefCondSpinTensor}
\end{equation}
 so only the purely spatial components of the particle's reference
spin tensor $S_{\refvalue}^{\mu\nu}$ are nonzero, 
\begin{equation}
S_{\refvalue}^{\mu\nu}=\begin{pmatrix}0 & 0 & 0 & 0\\
0 & 0 & S_{\refvalue,z} & -S_{\refvalue,y}\\
0 & -S_{\refvalue,z} & 0 & S_{\refvalue,x}\\
0 & S_{\refvalue,y} & -S_{\refvalue,x} & 0
\end{pmatrix}^{\mathclap{\mu\nu}},\label{eq:MassivePosEnergyRefSpinTensor}
\end{equation}
where the particle's spin three-vector $\vec S\defeq\parens{S^{yz},S^{zx},S^{xy}}$
was defined in \eqref{eq:DefSpinAngMom3Vec}. Thus, the invariant
quantity $s^{2}$ defined in \eqref{eq:4DSpinSquaredInvariantFrom3Vecs}
and characterizing the system's overall spin is non-negative: 
\begin{align}
s^{2} & =\vec S^{2}-\tilde{\vec S}^{2}\nonumber \\
 & =\vec S_{\refvalue}^{2}=S_{\refvalue,x}^{2}+S_{\refvalue,y}^{2}+S_{\refvalue,z}^{2}\geq0.\label{eq:MassivePosEnergySpinSquaredNonneg}
\end{align}

The corresponding reference value $W_{\refvalue}^{\mu}$ of the Pauli-Lubanski
pseudovector \eqref{eq:PauliLubanski4VecFromSpin} is then 
\begin{equation}
W_{\refvalue}^{\mu}=\parens{0,mc\,\vec S_{\refvalue}}^{\mu}.\label{eq:MassivePosEnergyRefPauliLubanski}
\end{equation}
 The Lorentz dot product of $W^{\mu}$ with itself therefore has the
non-negative, Lorentz-invariant value 
\begin{equation}
W^{2}\defeq w^{2}=m^{2}c^{2}s^{2}\geq0.\label{eq:MassivePosEnergySquarePauliLubanski}
\end{equation}

Notice that the reference value of the particle's dual spin three-vector
$\tilde{\vec S}\defeq\parens{S^{tx},S^{ty},S^{tz}}$, as defined in
\eqref{eq:DefDualSpinBoost3Vec}, vanishes in this case: 
\begin{equation}
\tilde{\vec S}_{\refvalue}=0.\label{eq:MassivePosEnergyDualSpin3VecZero}
\end{equation}
 It follows that the pseudoscalar invariant quantity $\tilde{s}^{2}$
defined in \eqref{eq:Def4DDualSpinSquaredInvariant} likewise vanishes:
\begin{equation}
\tilde{s}^{2}=\vec S\dotprod\tilde{\vec S}=\vec S_{\refvalue}\dotprod\tilde{\vec S}_{\refvalue}=0.\label{eq:MassivePosEnergyDualSpinSquareZero}
\end{equation}

On physical grounds, a localized system at fixed energy should have
a compact (that is, closed and bounded) set of states, because otherwise
its Boltzmann entropy under any equitable choice of coarse-graining
of the system's fixed-energy phase space would be infinite, and thus
the system would exhibit an infinite heat capacity.\footnote{For related arguments, see \citep{Wigner:1963iqmeom,Weinberg:1996tqtfi}.}
The compactness of a system's phase space at fixed energy in any one
inertial reference frame determines the compactness of the system's
phase space in any other inertial reference frame at the correspondingly
Lorentz-transformed energy, so it suffices to examine the compactness
of our particle's phase space at the fixed reference energy $E_{\refvalue}=p_{\refvalue}^{t}c=mc^{2}$
corresponding to the reference value \eqref{eq:DefMassivePosEnergyRef4Mom}
of the particle's four-momentum. The size of this subset of the particle's
phase space is determined by the set of all orthochronous Lorentz
transformations that leave the particle's reference four-momentum
$p_{\refvalue}^{\mu}\defeq\parens{mc,\vec 0}^{\mu}$ fixed. This
collection of transformations is called the little group of $p_{\refvalue}^{\mu}$.
In the present case, in which $p_{\refvalue}^{\mu}=\parens{mc,\vec 0}^{\mu}$,
this little group consists solely of the group $O\parens 3$ of three-dimensional
rotations and parity transformations, which collectively form a compact
set, so we are assured that the particle's phase space at any fixed
energy is likewise compact, as required.

To summarize, we see that a transitive group action of the orthochronous
Poincaré group for the case of a real and positive $m>0$ and positive
energy $E=p^{t}c>0$ describes a massive pointlike particle of inertial
mass $m$, non-negative spin-squared $s^{2}=\vec S_{\refvalue}^{2}\geq0$,
non-negative squared Pauli-Lubanski pseudovector $w^{2}=m^{2}c^{2}s^{2}\geq0$,
and timelike four-momentum $p^{\mu}=mu^{\mu}$. The particle moves
along a straight worldline in spacetime characterized by a normalized
four-velocity $u^{\mu}\defeq\dot{X}^{\mu}/\sqrt{-\dot{X}^{2}/c^{2}}$
and a three-dimensional velocity $\vec v=\vec pc^{2}/E$ that is always
slower than the speed of light, $\verts{\vec v}<c$, and the particle
has a compact phase space at any fixed value of its energy $E$.

\subsection{Massless, Positive-Energy Particles}

As our second example, we consider the case of $m=0$ and positive
energy $E=p^{t}c>0$. Because the system's four-momentum $p^{\mu}$
is therefore null, $p^{2}=0$, we again have from \eqref{eq:CausalPropertySignTemporalComponentTimelikeNull4Vecs}
that the condition $p^{t}>0$ is invariant under orthochronous Lorentz
transformations, and thus our positivity condition on $E$ is well-defined.

We can use $p^{2}=0$ to express the system's energy $E=p^{t}c$ in
terms of its three-dimensional momentum $\vec p$ as the mass-shell
relation 
\begin{equation}
E=\verts{\vec p}c.\label{eq:MasslessPosEnergyMassShellRelation}
\end{equation}

In contrast with the massive case, there is no rest frame in the massless
case\textemdash that is, we cannot set all three components of $\vec p$
to be zero, due to the pair of assumptions $p^{2}=0$ and $E=p^{t}c>0$.
There exist states in the system's phase space in which the four-momentum
$p^{\mu}$ has no $x$ or $y$ components, and we take one such value
of the four-momentum to be its reference value, for a fixed but arbitrarily
chosen value $E_{0}>0$ of the system's energy: 
\begin{equation}
p_{\refvalue}^{\mu}\defeq\parens{E_{\refvalue}/c,0,0,E_{\refvalue}/c}^{\mu}=\frac{E_{\refvalue}}{c}\parens{\delta_{t}^{\mu}+\delta_{t}^{z}}.\label{eq:DefMasslessPosEnergyRef4Mom}
\end{equation}

The positive-energy condition $E>0$ implies that the four-momentum
$p^{\mu}$ cannot vanish, and under our assumption of a strictly monotonic
parametrization $X^{\mu}\parens{\lambda}$, the four-velocity $\dot{X}^{\mu}$
also cannot vanish. With $m=0$, the relation \eqref{eq:FourMomDotProdFourVelRelationship}
degenerates to 
\begin{equation}
p\dotprod\dot{X}=0.\label{eq:MasslessPosEnergyFourMomDotProdFourVelRelationship}
\end{equation}
 A nonzero four-vector that has vanishing dot product with a null
four-vector must be parallel to that null four-vector. We can therefore
take the four-velocity $\dot{X}^{\mu}$ to be a null four-vector that
is parallel to the four-momentum $p^{\mu}$, 
\begin{equation}
p^{\mu}\propto\dot{X}^{\mu},\label{eq:MasslessPosEnergy4MomParallel4Vel}
\end{equation}
 which then ensures that the self-consistency condition \eqref{eq:DerivDualSquaredSpinTensorVanishesCondition},
$\dot{X}^{\mu}W_{\mu}=0$, is satisfied. 

The equation of motion \eqref{eq:4MomEOM}, $\dot{p}^{\mu}=0$, implies
that $p^{\mu}$ is constant along the system's worldline, so we can
always choose our parametrization $X^{\mu}\parens{\lambda}$ to make
the proportionality factor in \eqref{eq:MasslessPosEnergy4MomParallel4Vel}
equal to a constant: 
\begin{equation}
p^{\mu}=\parens{\textrm{const}}\dot{X}^{\mu}.\label{eq:MasslessPosEnergy4MomEqConstTimes4Vel}
\end{equation}
 We then have 
\begin{equation}
\ddot{X}^{\mu}=0,\label{eq:MasslessPosEnergy4VelConst}
\end{equation}
 so we see that the system describes a pointlike particle that travels
along a straight, null path in spacetime.

In addition, invoking the mass-shell relation \eqref{eq:MasslessPosEnergyMassShellRelation}
between the particle's energy $E$ and its three-dimensional momentum
$\vec p$, we see that the particle's three-dimensional velocity $\vec v$
is related to its three-dimensional momentum $\vec p$ according to
\begin{equation}
\vec v=\frac{d\vec X}{dt}=\frac{\dot{\vec X}}{\dot{T}}=\frac{\vec pc^{2}}{E}=\frac{\vec p}{\verts{\vec p}}c.\label{eq:MasslessPosEnergy3DMomVelRel}
\end{equation}
 Hence, the particle's speed $\verts{\vec v}$ is always equal to
the speed of light $c$: 
\begin{equation}
\verts{\vec v}=c.\label{eq:MasslessPosEnergySpeedEqSpeedOfLight}
\end{equation}

Turning to the particle's spin, we will find a much more nuanced story
than in the massive case.

The proportionality relationship $p^{\mu}\propto\dot{X}^{\mu}$ from
\eqref{eq:MasslessPosEnergy4MomParallel4Vel} together with the equation
of motion \eqref{eq:SpinTensorEOM} for the particle's spin tensor
$S^{\mu\nu}$ again imply that the particle's angular momentum \eqref{eq:Def4DOrbitalAngularMomentum}
and the particle's spin are separately conserved: 
\begin{align}
\dot{L}^{\mu\nu} & =\dot{X}^{\mu}p^{\nu}-\dot{X}^{\nu}p^{\mu}=0,\label{eq:MasslessPosEnergyOrbAngMomConst}\\
\dot{S}^{\mu\nu} & =0.\label{eq:MasslessPosEnergySpinConst}
\end{align}
 As in the massive case, the conservation law for $L^{ti}$ gives
back the formula \eqref{eq:MasslessPosEnergy3DMomVelRel} relating
the particle's three-dimensional velocity $\vec v$ to its three-dimensional
momentum $\vec p$.

However, the condition \eqref{eq:RefFourMomSpinTensorZero}, $p_{\refvalue,\mu}S_{\refvalue}^{\mu\nu}=0$,
is more complicated than it was in the massive case: 
\begin{equation}
-\frac{E_{\refvalue}}{c}\,S_{\refvalue}^{t\nu}+\frac{E_{\refvalue}}{c}\,S_{\refvalue}^{z\nu}=0.\label{eq:MasslessPosEnergyRefCondSpinTensor}
\end{equation}
 This equation implies that 
\begin{equation}
S_{\refvalue}^{t\nu}=S_{\refvalue}^{z\nu},\label{eq:MasslessPosEnergyRefEqualitySpinTensor}
\end{equation}
 or, equivalently, that the quantities 
\begin{equation}
\left.\begin{aligned}A & \defeq S_{x}+\tilde{S}_{y},\\
B & \defeq S_{y}-\tilde{S}_{x},
\end{aligned}
\hspace{1em}\right\} \label{eq:MasslessPosEnergyNoncompactLittleGroupGenerators}
\end{equation}
 and $\tilde{S}_{z}$ all vanish in the particle's reference state:
\begin{equation}
\left.\begin{aligned}A_{\refvalue} & \defeq S_{\refvalue,x}+\tilde{S}_{\refvalue,y}=0,\\
B_{\refvalue} & \defeq S_{\refvalue,y}-\tilde{S}_{\refvalue,x}=0,\\
\tilde{S}_{\refvalue,z} & =0.
\end{aligned}
\quad\right\} .\label{eq:MasslessPosEnergyNoncompactLittleGroupGeneratorsZero}
\end{equation}
  The reference value of the system's spin tensor is therefore 
\begin{equation}
S_{\refvalue}^{\mu\nu}=\begin{pmatrix}0 & S_{\refvalue,y} & -S_{\refvalue,x} & 0\\
-S_{\refvalue,y} & 0 & S_{\refvalue,z} & -S_{\refvalue,y}\\
S_{\refvalue,x} & -S_{\refvalue,z} & 0 & S_{\refvalue,x}\\
0 & S_{\refvalue,y} & -S_{\refvalue,x} & 0
\end{pmatrix}^{\mathclap{\mu\nu}}.\label{eq:MasslessPosEnergyRefSpinTensor}
\end{equation}
 In other words, the reference value of the particle's spin three-vector
$\vec S\defeq\parens{S^{yz},S^{zx},S^{xy}}$, as defined in \eqref{eq:DefSpinAngMom3Vec},
and the reference value of the particle's dual spin three-vector $\tilde{\vec S}\defeq\parens{S^{tx},S^{ty},S^{tz}}$,
as defined in \eqref{eq:DefDualSpinBoost3Vec}, are mutually perpendicular
and are related explicitly by 
\begin{equation}
\tilde{\vec S}_{\refvalue}=\vec S_{\refvalue}\crossprod\vec e_{z},\label{eq:MasslessPosEnergyDualSpin3VecFromSpin3Vec}
\end{equation}
 where $\vec e_{z}\defeq\parens{0,0,1}$ is the usual three-dimensional
Cartesian unit vector pointing along the positive $z$ axis.  It
follows that the pseudoscalar invariant quantity $\tilde{s}^{2}$
defined in \eqref{eq:Def4DDualSpinSquaredInvariant} vanishes, as
we also saw was true in the massive case: 
\begin{equation}
\tilde{s}^{2}=\vec S\dotprod\tilde{\vec S}=\vec S_{\refvalue}\dotprod\tilde{\vec S}_{\refvalue}=0.\label{eq:MasslessPosEnergyDualSpinSquaredZero}
\end{equation}

Meanwhile, the invariant quantity $s^{2}$ defined in \eqref{eq:4DSpinSquaredInvariantFrom3Vecs}
is non-negative, as in the massive case, but is now determined solely
by the $z$ component $S_{\refvalue,z}$ of the reference value $\vec S_{\refvalue}$
of the particle's spin three-vector: 
\begin{equation}
s^{2}=\vec S^{2}-\tilde{\vec S}^{2}=S_{\refvalue,z}^{2}\geq0.\label{eq:MasslessPosEnergySpinSquaredNonneg}
\end{equation}

In general, the projection of the particle's spin three-vector $\vec S$
onto the particle's three-dimensional momentum $\vec p\defeq\parens{p^{x},p^{y},p^{z}}$
is called the particle's helicity $\sigma$: 
\begin{equation}
\sigma\defeq\frac{\vec p}{\verts{\vec p}}\dotprod\vec S.\label{eq:DefHelicity}
\end{equation}
 The massless particle's helicity is insensitive to our reference
choice of energy $E_{\refvalue}$, due to the fact that $\vec p$
appears only as the ratio $\vec p/\verts{\vec p}$, and is also invariant
under proper rotations, due to the three-dimensional dot product.
Thus, $\sigma$ represents a fundamental feature of the particle in
the $m=0$ case that can only change under parity transformations
\eqref{eq:ParityTransf}, meaning that $\sigma$ is a pseudoscalar:
\begin{equation}
\sigma\mapsto-\sigma\quad\parens{\textrm{parity}}.\label{eq:HelicityParityTransf}
\end{equation}
 It follows from evaluating the definition \eqref{eq:DefHelicity}
of the helicity in the massless particle's reference state that the
helicity is equal to the $z$ component $S_{0,z}$ of the massless
particle's reference spin three-vector $\vec S_{0}$, up to a possible
sign that changes under parity transformations: 
\begin{equation}
\sigma=\left(\pm\right)^{\textrm{parity}}S_{0,z}.\label{eq:MasslessPosEnergyHelicityEqRefSpinZ}
\end{equation}
 We can therefore use $\sigma$ to write our expression \eqref{eq:MasslessPosEnergySpinSquaredNonneg}
for the invariant quantity $s^{2}$ as 
\begin{equation}
s^{2}=\sigma^{2}\geq0.\label{eq:MasslessPosEnergySpinSquaredFromHelicity}
\end{equation}

The reference value $W_{\refvalue}^{\mu}$ of the particle's Pauli-Lubanski
pseudovector \eqref{eq:PauliLubanski4VecFromSpin} is parallel to
the particle's reference four-momentum \eqref{eq:DefMasslessPosEnergyRef4Mom}:
\begin{equation}
W_{\refvalue}^{\mu}=\biggparens{S_{\refvalue,z}\frac{E_{\refvalue}}{c},0,0,S_{\refvalue,z}\frac{E_{\refvalue}}{c}}=S_{\refvalue,z}p_{\refvalue}^{\mu}.\label{eq:MasslessPosEnergyRefPauliLubanski}
\end{equation}
 More generally, $W^{\mu}$ is given in terms of the particle's helicity
\eqref{eq:DefHelicity} by 
\begin{equation}
W^{\mu}=\sigma p^{\mu},\label{eq:MasslessPosEnergyPauliLubanski}
\end{equation}
 where the pseudovector nature of $W^{\mu}$ is neatly captured by
the pseudoscalar nature of $\sigma$. As a consequence of the condition
$p^{2}=0$, we see that the invariant quantity $w^{2}$ defined in
\eqref{eq:SquarePauliLubanskiAsInvariant} vanishes: 
\begin{equation}
W^{2}\defeq w^{2}=0.\label{eq:MasslessPosEnergySquarePauliLubanski}
\end{equation}

As in the massive case, we will need to examine the compactness of
the subset of the particle's phase space at the fixed reference energy
$E_{\refvalue}=p_{\refvalue}^{t}c$. Again, this subspace is determined
by the little group of the particle's reference four-momentum \eqref{eq:DefMasslessPosEnergyRef4Mom},
meaning the set of all orthochronous Lorentz transformations that
leave $p_{\refvalue}^{\mu}\defeq\parens{E_{\refvalue}/c,0,0,E_{\refvalue}/c}^{\mu}$
invariant.

Let $\Lambda$ be a little-group transformation, so that $\Lambda p_{0}=p_{0}$.
For now, we will assume that $\Lambda$ does not involve a parity
transformation. As a trick for finding these little-group transformations,\footnote{See, for example, \citep{Weinberg:1996tqtfi}.}
let $v^{\mu}\defeq\parens{1,\vec 0}^{\mu}$ be a purely timelike four-vector.
Then 
\begin{align*}
\parens{\Lambda v}\dotprod p_{\refvalue} & =-\parens{\Lambda v}^{t}\frac{E_{\refvalue}}{c}+\parens{\Lambda v}^{z}\frac{E_{\refvalue}}{c}\\
\textrm{also} & =\parens{\Lambda v}\dotprod\parens{\Lambda p_{\refvalue}}=v\dotprod p_{\refvalue}=-\frac{E_{\refvalue}}{c}.
\end{align*}
 We conclude that 
\begin{equation}
\parens{\Lambda v}^{t}=1+\parens{\Lambda v}^{z},\label{eq:MasslessPosEnergyLittleGroupTransfOnUnitTimelike4VecIntermed}
\end{equation}
 and thus that $\parens{\Lambda v}^{\mu}$ has the form 
\begin{equation}
\parens{\Lambda v}^{\mu}=\parens{1+\zeta,\alpha,\beta,\zeta}^{\mu}\label{eq:MasslessPosEnergyLittleGroupTransfOnUnitTimelike4Vec}
\end{equation}
 for real-valued parameters $\alpha$, $\beta$, and $\zeta$. The
normalization condition $\parens{\Lambda v}^{2}=v^{2}=-1$ implies
that these three parameters are related by 
\begin{equation}
\zeta=\frac{\alpha^{2}+\beta^{2}}{2}.\label{eq:MasslessPosEnergyLittleGroupTransfOnUnitTimelike4VecParamRelation}
\end{equation}

The effect of the little-group Lorentz-transformation matrix $\Lambda$
on $v^{\mu}\defeq\parens{1,\vec 0}^{\mu}$ fixes $\Lambda$ up to
an overall three-dimensional rotation, and the little-group requirement
$\Lambda p_{\refvalue}=p_{\refvalue}$ further fixes $\Lambda$ up
to a rotation specifically around the $z$ axis. Hence, the most general
such proper orthochronous Lorentz-transformation matrix $\Lambda$
has the form 
\begin{equation}
\Lambda\parens{\alpha,\beta,\theta}=L\parens{\alpha,\beta}R\parens{\theta},\label{eq:MasslessPosEnergyLittleGroupMatrixFactorized}
\end{equation}
 where 
\begin{equation}
R\parens{\theta}\defeq\begin{pmatrix}1 & 0 & 0 & 0\\
0 & \cos\theta & \sin\theta & 0\\
0 & -\sin\theta & \cos\theta & 0\\
0 & 0 & 0 & 1
\end{pmatrix}\label{eq:MasslessPosEnergyLittleGroupMatrixRotationZ}
\end{equation}
 is a pure rotation by an angle $\theta$ around the $z$ axis, and
where 
\begin{equation}
L\parens{\alpha,\beta}\defeq\begin{pmatrix}1+\zeta & \alpha & \beta & -\zeta\\
\alpha & 1 & 0 & -\alpha\\
\beta & 0 & 1 & -\beta\\
\zeta & \alpha & \beta & 1-\zeta
\end{pmatrix}\label{eq:MasslessPosEnergyLittleGroupMatrixNoncompact}
\end{equation}
 is a complicated combination of proper orthochronous Lorentz boosts
and rotations satisfying the required condition $\Lambda^{\transp}\eta\Lambda=\eta$
from \eqref{eq:LorentzTransfsPreserveMinkMetric}. 

By straightforward calculations, one can show that 
\begin{align}
R\parens{\theta_{1}}R\parens{\theta_{2}} & =R\parens{\theta_{1}+\theta_{2}},\label{eq:MasslessPosEnergyRotationZCommutative}\\
L\parens{\alpha_{1},\beta_{1}}L\parens{\alpha_{2},\beta_{2}} & =L\parens{\alpha_{1}+\alpha_{2},\beta_{1}+\beta_{2}},\label{eq:MasslessPosEnergyNoncompactTransfCommutative}
\end{align}
 so rotations $R\parens{\theta}$ around the $z$ axis and the proper
orthochronous Lorentz transformations $L\parens{\alpha,\beta}$ respectively
form a pair of commutative subgroups of the particle's little group.
Furthermore, we have 
\begin{align}
 & R\parens{\theta}L\parens{\alpha,\beta}R^{-1}\parens{\theta}\nonumber \\
 & \quad=L\parens{\alpha\cos\theta+\beta\sin\theta,-\alpha\sin\theta+\beta\cos\theta},\label{eq:MasslessPosEnergyLittleGroupNoncomm}
\end{align}
 so we see that rotating $L\parens{\alpha,\beta}$ itself around the
$z$ axis has the effect of rotating the two-dimensional vector $\parens{\alpha,\beta}$.

The little group in this case is therefore the group $ISO\parens 2$
of translations and rotations in the two-dimensional Euclidean plane.
The subgroup $SO\parens 2$ consisting purely of rotations $R\parens{\theta}$
in the two-dimensional plane is compact, but the subgroup $\mathbb{R}^{2}$
consisting of two-dimensional translations $L\parens{\alpha,\beta}$
is noncompact. The consequence is that the particle's phase space
at the fixed reference four-momentum $p_{\refvalue}^{\mu}$ would
seem to be noncompact as well, leading to the thermodynamic problems
that we discussed earlier, as well as to various issues that arise
in the corresponding quantum field theory, such as those that are
explored in \citep{Abbott:1976mpcsi}, for example.\footnote{For a more optimistic alternative perspective, see \citep{SchusterToro:2013tcspwsfsa}.}

The particle's reference spacetime coordinates $X_{\refvalue}^{\mu}\defeq0$,
four-momentum $p_{\refvalue}^{\mu}\defeq\parens{E_{\refvalue}/c,0,0,E_{\refvalue}/c}^{\mu}$,
helicity $\sigma=\pm S_{\refvalue,z}$, and Pauli-Lubanski pseudovector
$W_{\refvalue}^{\mu}=\sigma p_{\refvalue}^{\mu}$ are all insensitive
to the noncompact transformations $L\parens{\alpha,\beta}$. However,
the particle's reference spin tensor \eqref{eq:MasslessPosEnergyRefSpinTensor}
transforms nontrivially under the action of $L\parens{\alpha,\beta}$:
\begin{align}
 & L\parens{\alpha,\beta}S_{\refvalue}L^{\transp}\parens{\alpha,\beta}\nonumber \\
 & =S_{\refvalue}+\begin{pmatrix}0 & -\beta S_{\refvalue,z} & \alpha S_{\refvalue,z} & 0\\
\beta S_{\refvalue,z} & 0 & 0 & \beta S_{\refvalue,z}\\
\alpha S_{\refvalue,z} & 0 & 0 & -\alpha S_{\refvalue,z}\\
0 & -\beta S_{\refvalue,z} & \alpha S_{\refvalue,z} & 0
\end{pmatrix}.\label{eq:MasslessPosEnergyNoncompactTransfSpinTensor}
\end{align}
 In particular, at the level of the massless particle's reference
spin three-vector $\vec S_{0}$, we have the transformation 
\begin{equation}
\vec S_{0}\mapsto\vec S_{0}+\parens{-\alpha S_{0,z},-\beta S_{0,z},0}.\label{eq:MasslessPosEnergyNoncompactTransfRefSpin3Vec}
\end{equation}
Notice that the discrepant spin components $\parens{-\alpha S_{0,z},-\beta S_{0,z},0}$
here are perpendicular to the particle's reference three-momentum
$\vec p_{\refvalue}=\parens{0,0,E_{\refvalue}/c}$, and, furthermore,
parametrize \emph{all} possible values of components perpendicular
to $\vec p_{0}$, contingent on the $z$ component of $\vec S_{0}$
being nonzero, $S_{0,z}\ne0$. The fact that the discrepant spin components
are perpendicular to $\vec p_{0}$ is ultimately guaranteed by the
invariance of the massless particle's helicity $\sigma\equiv\parens{\vec p/\verts{\vec p}}\dotprod\vec S$
under all proper Lorentz transformations, together with the invariance
of the particle's reference momentum $\vec p_{0}$ under little-group
transformations.

Moreover, suppose that two states of our massless particle both have
four-momentum equal to the reference four-momentum $p_{0}^{\mu}=\parens{E/c,0,0,E/c}$,
but have different values of the spin tensor $S_{0}^{\mu\nu}\ne S^{\prime\mu\nu}$.
The invariance of the helicity $\sigma\defeq\parens{\vec p/\verts{\vec p}}\dotprod\vec S$
over the entire phase space, up to a possible minus sign under parity
transformations, then implies that 
\begin{equation}
S_{0,z}^{\prime}=\pm S_{0,z}.\label{eq:MasslessPosEnergySpinZDoubleValued}
\end{equation}
 Hence, all states with the reference momentum $p_{0}^{\mu}$ have
spin-$z$ component equal either to $S_{0,z}$ or to $-S_{0,z}$,
where these two sets of states have opposite helicity $\sigma$. Fixing
the helicity, such states can then differ at most in their spin-$x$
and spin-$y$ components.

The real-valued parameters $\alpha$ and $\beta$ appearing in the
little-group transformations \eqref{eq:MasslessPosEnergyNoncompactTransfSpinTensor}
and \eqref{eq:MasslessPosEnergyNoncompactTransfRefSpin3Vec} are arbitrary,
and parametrize a noncompact set of states at fixed reference momentum
$p_{0}^{\mu}$. As a consequence, the only way to ensure that the
massless particle's phase space at the fixed reference energy $E=p_{\refvalue}^{t}c$
is compact is to institute an equivalence relation in which we declare
that any two states of the form $\parens{X_{\refvalue},p_{\refvalue},S_{\refvalue}}$
and $\parens{X_{\refvalue},p_{\refvalue},S^{\prime}}$ that have the
same helicity $\sigma$ and that differ solely in their spin components
are to be regarded as the \emph{same} physical state: 
\begin{equation}
\parens{X_{\refvalue},p_{\refvalue},S_{\refvalue}}\cong\parens{X_{\refvalue},p_{\refvalue},S^{\prime}}\quad\left[\textrm{for fixed }\sigma\right].\label{eq:MasslessPosEnergyEquivRelationGaugeInvarianceRef}
\end{equation}
 This equivalence relation immediately generalizes to arbitrary states
as 
\begin{equation}
\parens{X,p,S}\cong\parens{X,p,S^{\prime}}\quad\left[\textrm{for fixed }\sigma\right],\label{eq:MasslessPosEnergyEquivRelationGaugeInvarianceGeneral}
\end{equation}
 where the two states have the same spacetime coordinates $X^{\mu}$,
four-momentum $p^{\mu}$, and helicity $\sigma$.

The equivalence relation \eqref{eq:MasslessPosEnergyEquivRelationGaugeInvarianceGeneral}
is a new result, and is another important example of a gauge invariance,
distinct from the reparametrization invariance of a manifestly covariant
action functional that we introduced earlier. A space with an equivalence
relation is known as a quotient space, and so we see that the phase
space of a massless $m=0$ particle with nonzero spin $s^{2}\ne0$
is a quotient space under the gauge invariance \eqref{eq:MasslessPosEnergyEquivRelationGaugeInvarianceGeneral}.

All physical observables must therefore be gauge invariant, as is
indeed the case for the particle's spacetime coordinates $X^{\mu}$,
its four-momentum $p^{\mu}$, its helicity $\sigma$, and its Pauli-Lubanski
pseudovector $W^{\mu}=\sigma p^{\mu}$. By contrast, the equivalence
relation \eqref{eq:MasslessPosEnergyEquivRelationGaugeInvarianceGeneral},
together with the invariance of the helicity $\sigma$, as defined
in \eqref{eq:DefHelicity}, implies that 
\begin{equation}
\sigma\defeq\frac{\vec p}{\verts{\vec p}}\dotprod\vec S\cong\frac{\vec p}{\verts{\vec p}}\dotprod\vec S^{\prime},\label{eq:MasslessPosEnergyInvHelicityGaugeEquiv}
\end{equation}
 so components of the particle's spin tensor $S^{\mu\nu}$ that are
perpendicular to the particle's three-momentum $\vec p$\textemdash such
as $S^{yz}=S_{x}$ and $S^{zx}=S_{y}$ if $\vec p$ points along the
$z$ direction\textemdash are not gauge invariant, and are consequently
not physical observables.

Combining the equivalence relation \eqref{eq:MasslessPosEnergyEquivRelationGaugeInvarianceGeneral}
with the relationship $\sigma=\pm S_{0,z}$ from \eqref{eq:MasslessPosEnergyHelicityEqRefSpinZ}
and our condition \eqref{eq:MasslessPosEnergySpinZDoubleValued},
we see that a massless particle has just \emph{two} physical spin
states for each fixed value of the particle's three-momentum $\vec p$.
These two physical spin states corresponding to the two parity-related
helicities $\pm\sigma$.

Summarizing our results, we see that a transitive group action of
the orthochronous Poincaré group with $m=0$ and positive energy $E=p^{t}c>0$
describes the phase space of a massless particle with null four-momentum
$p^{\mu}$, helicity $\sigma$, non-negative spin-squared $s^{2}=\sigma^{2}\geq0$,
and a null Pauli-Lubanski pseudovector $W^{\mu}=\sigma p^{\mu}$.
The particle moves at the speed of light $c$ along a null worldline
in spacetime with null four-velocity $\dot{X}^{\mu}$, and the particle's
spin tensor $S^{\mu\nu}$ is uniquely defined only up to gauge transformations
$S^{\mu\nu}\mapsto S^{\prime\mu\nu}$ for which $S^{\prime\mu\nu}$
differs from $S^{\mu\nu}$ solely by components perpendicular to the
particle's three-momentum $\vec p$. This gauge invariance implies
that the massless particle has just two physical spin states for each
fixed value of $\vec p$, corresponding to the two helicities $\pm\sigma$,
which are related to each other by parity transformations.

As an aside, we note that in the counterpart quantum theory, spin
components that are \emph{perpendicular }to the particle's direction
of motion correspond to linear polarizations that are \emph{longitudinal},
meaning that they are \emph{parallel} to the particle's direction
of motion. Accordingly, spin components that are \emph{parallel} to
the particle's direction of motion correspond to \emph{transverse}
linear polarizations. So in the quantum version of this story, gauge-invariant
observables are those that are insensitive to the particle's longitudinal
linear polarizations.

Moving on to connections with classical field theory, recall our earlier
analysis of the relationship between the classical-particle condition
$p_{\mu}S^{\mu\nu}=0$ from \eqref{eq:FourMomSpinTensorZeroPhysicalCondition}
and the Lorenz equation $\partial_{\mu}A^{\mu}=0$ from \eqref{eq:LorenzEquation}
for a spin-1 field theory. In particular, we showed that the Lorenz
equation $\partial_{\mu}A^{\mu}=0$, applied to the case \eqref{eq:Spin1FieldPlaneWave}
of a monochromatic plane wave $A^{\mu}=\varepsilon^{\mu}\exp\parens{ip\dotprod x/\hbar}$
with polarization four-vector $\varepsilon^{\mu}$, yielded the condition
\eqref{eq:MomentumLorenzEquation}, $p_{\mu}\varepsilon^{\mu}=0$,
which was precisely analogous to $p_{\mu}S^{\mu\nu}=0$, and played
the same role of eliminating unphysical spin states. 

Similarly, the classical-particle gauge invariance \eqref{eq:MasslessPosEnergyEquivRelationGaugeInvarianceGeneral}
is directly analogous to electromagnetic gauge invariance: 
\begin{equation}
A_{\mu}\cong A_{\mu}+\partial_{\mu}f.\label{eq:Spin1FieldGaugeInvariance}
\end{equation}
 Indeed, for the plane-wave case $A^{\mu}=\varepsilon^{\mu}\exp\parens{ip\dotprod x/\hbar}$,
with $f=\alpha\exp\parens{ip\dotprod x/\hbar}$, the identification
\eqref{eq:Spin1FieldGaugeInvariance} yields the condition that 
\begin{equation}
\varepsilon^{\mu}\cong\varepsilon^{\mu}+\parens{i\alpha/\hbar}p^{\mu},\label{eq:Spin1FieldMomentumGaugeInvariance}
\end{equation}
 which is likewise responsible for eliminating unphysical spin states.

Both our classical-particle gauge invariance \eqref{eq:MasslessPosEnergyEquivRelationGaugeInvarianceGeneral}
and electromagnetic gauge invariance \eqref{eq:Spin1FieldGaugeInvariance}
have nontrivial implications for the allowed form of interactions
between systems, as any such interactions must be insensitive to quantities
that are not gauge invariant. For the case of electromagnetism, the
gauge potential $A_{\mu}$ cannot directly appear in Lorentz-covariant
field equations, but can only appear indirectly through the gauge-invariant
Faraday tensor $F_{\mu\nu}\defeq\partial_{\mu}A_{\nu}-\partial_{\nu}A_{\mu}$.
In an analogous way, for our classical massless particle, the spin
tensor $S^{\mu\nu}$ cannot directly appear in Lorentz-covariant interaction
terms in equations of motion that couple the particle to other systems. 

Interaction terms involving the particle's four-momentum $p^{\mu}$
or Pauli-Lubanski pseudovector $W^{\mu}=\sigma p^{\mu}$ would both
be permitted, although they get weak for small momentum, corresponding
in quantum mechanics to large distances. We therefore anticipate that
massless particles with classically large total spin $s\gg\hbar$
cannot mediate long-range interactions, and, indeed, a quantum version
of our classification of particle-types suggests that long-range interactions
are mediated only by massless particles with total spin less than
or equal to $2\hbar$.\footnote{Again, for an alternative point of view, see \citep{SchusterToro:2013tcspwsfsa}.}

\subsection{The Massless Limit}

It is an enlightening exercise to re-examine the massless case $m=0$
from the perspective of the massive case $m>0$ in the limit $m\to0$.
Along the way, we will provide a deeper explanation for the emergence
of gauge invariance, as well as derive a classical-particle version
of the Higgs mechanism.

To start, notice that our original choice \eqref{eq:DefMassivePosEnergyRef4Mom}
of reference four-momentum in the massive case, $p_{\refvalue}^{\mu}\defeq\parens{mc,\vec 0}^{\mu}$,
does not have an appropriate massless limit. However, our choice of
reference four-momentum is entirely arbitrary apart from the condition
that $p^{2}=-m^{2}c^{2}$ from \eqref{eq:Def4DMassSquaredAsInvariant},
so we can instead choose it to be 
\begin{align}
\bar{p}^{\mu} & \defeq\parens{\bar{p}^{t},0,0,\bar{p}^{z}}^{\mu}\nonumber \\
 & =\parens{\sqrt{\parens{\bar{p}^{z}}^{2}+m^{2}c^{2}},0,0,\bar{p}^{z}}^{\mu}.\label{eq:DefMassivePosEnergyRef4MomAlt}
\end{align}
 The bars in this notation are solely meant to distinguish this choice
of reference momentum from our original choice $p_{\refvalue}^{\mu}\defeq\parens{mc,\vec 0}^{\mu}$
in \eqref{eq:DefMassivePosEnergyRef4Mom}.

The massless limit $m\to0$ of this alternative reference four-momentum
replicates the reference four-momentum \eqref{eq:DefMasslessPosEnergyRef4Mom}
that we chose for the case of a massless particle: 
\begin{equation}
\lim_{m\to0}\bar{p}^{\mu}=\parens{E_{\refvalue}/c,0,0,E_{\refvalue}/c}^{\mu},\quad E_{\refvalue}\defeq\bar{p}^{z}c.\label{eq:MassivePosEnergyMasslessLimitFourMom}
\end{equation}
 Moreover, the choice \eqref{eq:DefMassivePosEnergyRef4MomAlt} is
related to our original reference four-momentum \eqref{eq:DefMassivePosEnergyRef4Mom},
\[
p_{\refvalue}^{\mu}=\parens{mc,\vec 0}^{\mu},
\]
 by a simple Lorentz boost $\bar{\Lambda}$ along the $z$ direction,
\begin{equation}
\bar{p}^{\mu}=\tud{\bar{\Lambda}}{\mu}{\nu}p_{\refvalue}^{\nu},\label{eq:MassivePosEnergyFourMomAltFromLorentzBoost}
\end{equation}
 where 
\begin{equation}
\bar{\Lambda}\defeq\begin{pmatrix}{\displaystyle \frac{\bar{p}^{t}}{mc}} & 0 & 0 & {\displaystyle \frac{\bar{p}^{z}}{mc}}\\
0 & 1 & 0 & 0\\
0 & 0 & 1 & 0\\
{\displaystyle \frac{\bar{p}^{z}}{mc}} & 0 & 0 & {\displaystyle \frac{\bar{p}^{t}}{mc}}
\end{pmatrix}.\label{eq:MassivePosEnergyRefLorentzBoost}
\end{equation}

It follows that the new reference value $\bar{S}^{\mu\nu}$ of the
massive particle's spin tensor is related to its old reference value
$S_{\refvalue}^{\mu\nu}$ from \eqref{eq:MassivePosEnergyRefSpinTensor}
according to 
\begin{align}
 & \bar{S}^{\mu\nu}\defeq\parens{\bar{\Lambda}S_{\refvalue}\bar{\Lambda}^{\transp}}^{\mu\nu}\nonumber \\
 & =\begin{pmatrix}0 & {\displaystyle \frac{\bar{p}^{z}}{mc}}S_{\refvalue,y} & -{\displaystyle \frac{\bar{p}^{z}}{mc}}S_{\refvalue,x} & 0\\
-{\displaystyle \frac{\bar{p}^{z}}{mc}}S_{\refvalue,y} & 0 & S_{\refvalue,z} & -{\displaystyle \frac{\bar{p}^{t}}{mc}}S_{\refvalue,y}\\
{\displaystyle \frac{\bar{p}^{z}}{mc}}S_{\refvalue,x} & -S_{\refvalue,z} & 0 & {\displaystyle \frac{\bar{p}^{t}}{mc}}S_{\refvalue,x}\\
0 & {\displaystyle \frac{\bar{p}^{t}}{mc}}S_{\refvalue,y} & -{\displaystyle \frac{\bar{p}^{t}}{mc}}S_{\refvalue,x} & 0
\end{pmatrix}^{\mathclap{\mu\nu}}.\label{eq:MassivePosEnergyNewRefSpinTensorFromBoostOld}
\end{align}
 Both $\bar{p}^{t}$ and $\bar{p}^{z}$ approach the finite, nonzero
value $E_{\refvalue}/c>0$ in the massless limit $m\to0$, so the
components of $\bar{S}^{\mu\nu}$ that involve factors of $\bar{p}^{t}/mc$
or $\bar{p}^{z}/mc$ diverge in that limit. Furthermore, the particle's
spin-squared scalar $s^{2}$ continues to have its invariant value
\eqref{eq:MassivePosEnergySpinSquaredNonneg}, which, despite remaining
well-defined in the limit $m\to0$, does not end up agreeing with
the corresponding massless particle's spin-squared scalar \eqref{eq:MasslessPosEnergySpinSquaredNonneg}:
\begin{align}
s^{2}=S_{\refvalue,x}^{2}+S_{\refvalue,y}^{2} & +S_{\refvalue,z}^{2}\quad\parens{\textrm{massive}}\nonumber \\
 & \ne S_{\refvalue,z}^{2}\quad\parens{\textrm{massless}}.\label{eq:MassivePosEnergySpinSqDisagreeWithMassless}
\end{align}

Meanwhile, the new reference value $\bar{W}^{\mu}$ of the particle's
Pauli-Lubanski pseudovector is related to its old reference value
$W_{\refvalue}^{\mu}\defeq\parens{0,mc\,\vec S_{\refvalue}}^{\mu}$
from \eqref{eq:MassivePosEnergyRefPauliLubanski} according to 
\begin{align}
\bar{W}^{\mu} & =\tud{\bar{\Lambda}}{\mu}{\nu}W_{\refvalue}^{\mu}\nonumber \\
 & =\parens{\bar{p}^{z}\,S_{\refvalue,z},mc\,S_{\refvalue,x},mc\,S_{\refvalue,y},\bar{p}^{t}\,S_{\refvalue,z}}^{\mu}.\label{eq:MassivePosEnergyRefPauliLubanskiAlt}
\end{align}
 This expression has a well-defined massless limit that precisely
agrees with the reference value \eqref{eq:MasslessPosEnergyRefPauliLubanski}
of the Pauli-Lubanski pseudovector for a massless particle: 
\begin{equation}
\lim_{m\to0}\bar{W}^{\mu}=\biggparens{S_{\refvalue,z}\frac{E_{\refvalue}}{c},0,0,S_{\refvalue,z}\frac{E_{\refvalue}}{c}}^{\mu}.\label{eq:MassivePosEnergyPauliLubanskiAltMasslessLimit}
\end{equation}
 To make contact with the massless case, we can therefore focus our
efforts on the spin tensor \eqref{eq:MassivePosEnergyNewRefSpinTensorFromBoostOld}.

An important hint is the discrete discrepancy \eqref{eq:MassivePosEnergySpinSqDisagreeWithMassless}
between the spin-squared scalar $s^{2}$ in the massive and massless
cases, signaling that the massive case features spin degrees of freedom
that need to be removed before taking the massless limit. As we will
see, removing these extraneous spin degrees of freedom will require
formally enlarging our massive particle's phase space while simultaneously
introducing a compensating equivalence relation to ensure that we
are not adding any physically new states to the system. This approach
corresponds to an analogous construction in quantum field theory whose
origins go back to the work of Stueckelberg in \citep{Stueckelberg:1938dwefk}.

At the conclusion of this procedure, we will be able to isolate and
eliminate the extraneous spin degrees of freedom. Moreover, we will
find that the equivalence relation that we introduced along the way
becomes the gauge invariance \eqref{eq:MasslessPosEnergyEquivRelationGaugeInvarianceGeneral}
in the massless limit.

We begin by redefining the $x$ and $y$ components of the reference
value $\bar{\vec S}=\parens{\bar{S}_{x},\bar{S}_{y},\bar{S}_{z}}$
of the massive particle's spin three-vector according to 
\begin{equation}
\begin{pmatrix}\bar{S}_{x}\\
\bar{S}_{y}
\end{pmatrix}\mapsto\frac{mc}{\bar{p}^{t}}\begin{pmatrix}\bar{S}_{x}+\bar{p}^{t}\varphi_{x}\\
\bar{S}_{y}+\bar{p}^{t}\varphi_{y}
\end{pmatrix}=\frac{mc}{\bar{p}^{t}}\begin{pmatrix}\bar{S}_{x}\\
\bar{S}_{y}
\end{pmatrix}+mc\begin{pmatrix}\varphi_{x}\\
\varphi_{y}
\end{pmatrix},\label{eq:MassivePosEnergyStueckelbergRedefSpinTensor}
\end{equation}
 where $\varphi_{x}\parens{\lambda}$ and $\varphi_{y}\parens{\lambda}$
are arbitrary new functions on the particle's worldline. The particle's
spin tensor \eqref{eq:MassivePosEnergyNewRefSpinTensorFromBoostOld}
then takes the form 
\begin{align}
\bar{S}^{\mu\nu} & =\begin{pmatrix}0 & {\displaystyle \frac{\bar{p}^{z}}{\bar{p}^{t}}}S_{\refvalue,y} & -{\displaystyle \frac{\bar{p}^{z}}{\bar{p}^{t}}}S_{\refvalue,x} & 0\\
-{\displaystyle \frac{\bar{p}^{z}}{\bar{p}^{t}}}S_{\refvalue,y} & 0 & S_{\refvalue,z} & -S_{\refvalue,y}\\
{\displaystyle \frac{\bar{p}^{z}}{\bar{p}^{t}}}S_{\refvalue,x} & -S_{\refvalue,z} & 0 & S_{\refvalue,x}\\
0 & S_{\refvalue,y} & -S_{\refvalue,x} & 0
\end{pmatrix}^{\mathclap{\mu\nu}}\nonumber \\
 & +\begin{pmatrix}0 & \bar{p}^{z}\varphi_{y} & -\bar{p}^{z}\varphi_{x} & 0\\
-\bar{p}^{z}\varphi_{y} & 0 & 0 & -\bar{p}^{t}\varphi_{y}\\
\bar{p}^{z}\varphi_{x} & 0 & 0 & \bar{p}^{t}\varphi_{x}\\
0 & \bar{p}^{t}\varphi_{y} & -\bar{p}^{t}\varphi_{x} & 0
\end{pmatrix}^{\mathclap{\mu\nu}},\label{eq:MassivePosEnergyRedefinedSpinTensor}
\end{align}
 where we have chosen the various factors of $m$, $c$, $\bar{p}^{t}$,
and $\bar{p}^{z}$ in the redefinition \eqref{eq:MassivePosEnergyStueckelbergRedefSpinTensor}
to ensure that the two tensors appearing in \eqref{eq:MassivePosEnergyRedefinedSpinTensor}
separately satisfy the fundamental condition $\bar{p}_{\mu}\parens{\cdots}^{\mu\nu}=0$
from \eqref{eq:FourMomSpinTensorZeroPhysicalCondition}. The particle's
spin-squared scalar $s^{2}$, as originally defined in \eqref{eq:Def4DSpinSquaredAsInvariant},
now becomes 
\begin{align}
s^{2} & =\biggparens{1-\biggparens{\frac{\bar{p}^{z}}{\bar{p}^{t}}}^{2}}\Big(\parens{S_{\refvalue,x}+\bar{p}^{t}\varphi_{x}}^{2}\nonumber \\
 & \qquad\qquad\qquad\qquad+\parens{S_{\refvalue,y}+\bar{p}^{t}\varphi_{y}}^{2}\Big)+S_{\refvalue,z}^{2}.\label{eq:MassivePosEnergyRedefinedSpinSqScalar}
\end{align}

Notice that the particle's spin tensor \eqref{eq:MassivePosEnergyRedefinedSpinTensor}
is invariant under the simultaneous transformations 
\begin{align}
\begin{pmatrix}\bar{S}_{x}\\
\bar{S}_{y}
\end{pmatrix} & \mapsto\begin{pmatrix}\bar{S}_{x}\\
\bar{S}_{y}
\end{pmatrix}-\bar{p}^{t}\begin{pmatrix}f_{x}\\
f_{y}
\end{pmatrix},\label{eq:MassivePosEnergyGaugeTransfPerpSpin}\\
\begin{pmatrix}\varphi_{x}\\
\varphi_{y}
\end{pmatrix} & \mapsto\begin{pmatrix}\varphi_{x}\\
\varphi_{y}
\end{pmatrix}+\begin{pmatrix}f_{x}\\
f_{y}
\end{pmatrix},\label{eq:MassivePosEnergyGaugeTransfAncillary}
\end{align}
 where $f_{x}\parens{\lambda},f_{y}\parens{\lambda}$ are arbitrary
functions on the particle's worldline.  We claim that our massive
particle's original phase space, with states denoted by $\parens{X,p,S}$,
is equivalent to a \emph{formally enlarged} phase space consisting
of states 
\begin{equation}
\parens{X,p,S,\varphi}\label{eq:MassivePosEnergyFormallyEnlargeStates}
\end{equation}
 under the equivalence relation 
\begin{equation}
\parens{\bar{X},\bar{p},\bar{S},\varphi}\cong\parens{\bar{X},\bar{p},\bar{S}-\bar{p}^{t}f,\varphi+f},\label{eq:MassivePosEnergyGaugeEquivalenceRelation}
\end{equation}
 suitably generalized from the reference state $\parens{\bar{X},\bar{p},\bar{S},\varphi}$
to general states $\parens{X,p,S,\varphi}$ of the system. To see
why, observe that the specific choice 
\begin{equation}
f\defeq-\varphi,\label{eq:MassivePosEnergyFixGaugeSuccinct}
\end{equation}
 or, more explicitly, 
\begin{equation}
\begin{pmatrix}f_{x}\\
f_{y}
\end{pmatrix}\defeq-\begin{pmatrix}\varphi_{x}\\
\varphi_{y}
\end{pmatrix}\label{eq:MassivePosEnergyFixGauge}
\end{equation}
 makes clear that the state $\parens{\bar{X},\bar{p},\bar{S},\varphi}$
is equivalent to the state $\parens{\bar{X},\bar{p},\bar{S}+\bar{p}^{t}\varphi,0}$,
which gives us back the state $\parens{\bar{X},\bar{p},\bar{S}}$
after undoing the redefinition \eqref{eq:MassivePosEnergyStueckelbergRedefSpinTensor}
of $\bar{S}^{\mu\nu}$.

The system's redefined spin tensor \eqref{eq:MassivePosEnergyRedefinedSpinTensor}
now has a nice massless limit, 
\begin{align}
\lim_{m\to0}\bar{S}^{\mu\nu} & =\begin{pmatrix}0 & S_{\refvalue,y} & -S_{\refvalue,x} & 0\\
-S_{\refvalue,y} & 0 & S_{\refvalue,z} & -S_{\refvalue,y}\\
S_{\refvalue,x} & -S_{\refvalue,z} & 0 & S_{\refvalue,x}\\
0 & S_{\refvalue,y} & -S_{\refvalue,x} & 0
\end{pmatrix}^{\mathclap{\mu\nu}}\nonumber \\
 & +\frac{E}{c}\begin{pmatrix}0 & \varphi_{y} & -\varphi_{x} & 0\\
-\varphi_{y} & 0 & 0 & -\varphi_{y}\\
\varphi_{x} & 0 & 0 & \varphi_{x}\\
0 & \varphi_{y} & -\varphi_{x} & 0
\end{pmatrix}^{\mathclap{\mu\nu}},\label{eq:MassivePosEnergyRedefinedSpinTensorMasslessLimit}
\end{align}
 as does the particle's spin-squared scalar \eqref{eq:MassivePosEnergyRedefinedSpinSqScalar},
\begin{equation}
\lim_{m\to0}s^{2}=S_{\refvalue,z}^{2}.\label{eq:MassivePosEnergyRedefinedSpinSqMasslessLimit}
\end{equation}

Our system fundamentally has the same number of degrees of freedom
as it had before we took the massless limit, but we see that the degrees
of freedom describing spin components perpendicular to the particle's
reference three-momentum $\bar{\vec p}$ no longer contribute to the
particle's spin-squared scalar $s^{2}$, which agrees with the spin-squared
scalar \eqref{eq:MasslessPosEnergySpinSquaredNonneg} of the massless
case. If we now formally remove the spin degrees of freedom $\varphi_{x},\varphi_{y}$
by setting them equal to zero, then the particle's spin tensor \eqref{eq:MassivePosEnergyRedefinedSpinTensorMasslessLimit}
reduces to the reference value of the massless spin tensor \eqref{eq:MasslessPosEnergyRefSpinTensor},
and our equivalence relation \eqref{eq:MassivePosEnergyGaugeEquivalenceRelation}
reduces to the gauge invariance \eqref{eq:MasslessPosEnergyEquivRelationGaugeInvarianceRef}.

Notice that if we run all the arguments of this section in reverse,
then we can convert a massless particle with spin into a massive particle
by introducing additional spin degrees of freedom. We therefore obtain
a classical version of the celebrated Higgs mechanism.

To see the connection with the field-theoretic Higgs mechanism in
more detail, recall the case of a massless spin-1 field, whose corresponding
quantum-mechanical boson has two physical spin states. If we spontaneously
break the gauge symmetry, then the spin-1 field gains a positive mass
along with an additional spin state that is acquired from the Higgs
field, so that the corresponding quantum-mechanical boson ends up
with the correct three physical spin states for a massive spin-1 particle.

In a similar way, suppose that we start with a massless classical
particle with nonzero spin. From our preceding work, we know that
the particle has precisely two physical spin states, corresponding
to the two helicities $\pm\sigma$. Reversing the logic of this section,
we can convert our massless particle into a massive particle by augmenting
the particle with spin states from the ``Higgs'' degrees of freedom
$\varphi_{x},\varphi_{y}$.

\subsection{Tachyons}

The case $m^{2}<0$ is also interesting. The invariant quantity $m$
is now purely imaginary and is therefore of the form $m=i\mu$ for
a real constant $\mu$. The system's four-momentum $p^{\mu}$ is spacelike,
$p^{2}=\mu^{2}c^{2}>0$, so its temporal component $p^{t}$ does not
have a definite sign under orthochronous Lorentz transformations.
As a consequence, we cannot impose a positivity condition on the system's
energy.

We can use $p^{2}=\mu^{2}c^{2}$ to express the system's energy $E=p^{t}c$
in terms of its three-dimensional momentum $\vec p$ as the mass-shell
relation 
\begin{equation}
E=\sqrt{\vec p^{2}c^{2}-\mu^{2}c^{2}}.\label{eq:TachyonMassShellRelation}
\end{equation}
 For convenience, we will take the system's reference four-momentum
to be purely spacelike and aligned with the $z$ direction: 
\begin{equation}
p_{\refvalue}^{\mu}\defeq\parens{0,0,0,\mu c}^{\mu}=\mu c\,\delta_{z}^{\mu}.\label{eq:TachyonRef4Mom}
\end{equation}

Once again, the four-momentum $p^{\mu}$ and the four-velocity $\dot{X}^{\mu}$
are non-vanishing, and so the relation \eqref{eq:FourMomFourVelSqrRootRelationship},
\[
m\sqrt{-\dot{X}^{2}/c^{2}}\,p^{\mu}=\mp m^{2}\dot{X}^{\mu},
\]
 becomes 
\begin{equation}
\sqrt{-\dot{X}^{2}/c^{2}}\,p^{\mu}=\mp i\mu\dot{X}^{\mu}.\label{eq:FourMomFourVelSqrRootRelationshipTachyon}
\end{equation}
 Because the right-hand side is imaginary, this equality implies that
$\dot{X}^{2}>0$, so the four-velocity $\dot{X}^{\mu}$ is likewise
spacelike and is related to the four-momentum $p^{\mu}$ by 
\begin{equation}
p^{\mu}=\mu\frac{\dot{X}^{\mu}}{\sqrt{\dot{X}^{2}/c^{2}}},\label{eq:Tachyon4Mom4VelParallel}
\end{equation}
 where we have taken the positive sign by assuming that our parametrization
$X^{\mu}\parens{\lambda}$ points in the positive direction along
$p^{\mu}$. This relation between $p^{\mu}$ and $\dot{X}^{\mu}$
again ensures that the self-consistency condition \eqref{eq:DerivDualSquaredSpinTensorVanishesCondition},
$\dot{X}^{\mu}W_{\mu}=0$, is satisfied.

The equation of motion \eqref{eq:4MomEOM} for the system's four-momentum,
$\dot{p}^{\mu}=0$, then tells us that the system's path has a fixed,
spacelike direction in spacetime. A calculation of the system's three-dimensional
velocity $\vec v$ using the mass-shell relation \eqref{eq:TachyonMassShellRelation}
yields the result 
\begin{equation}
\vec v=\frac{d\vec X}{dt}=\frac{\dot{\vec X}}{\dot{T}}=\frac{\vec pc^{2}}{E}=\frac{\vec p}{\verts{\vec p}}\frac{c}{\sqrt{1-\mu^{2}c^{2}/\vec p^{2}}}.\label{eq:Tachyon3DVelMomRelation}
\end{equation}
 Hence, the system's speed $\verts{\vec{\vec v}}$ is always \emph{greater}
than the speed of light $c$: 
\begin{equation}
\verts{\vec v}>c.\label{eq:TachyonSpeedFasterThanSpeedOfLight}
\end{equation}
 Such a system is appropriately called a tachyon, from the Greek
for ``swift.'' 

By the same reasoning as in the massive and massless cases, a tachyon's
orbital and spin angular momenta are separately conserved, 
\begin{align}
\dot{L}^{\mu\nu} & =0,\label{eq:TachyonOrbAngMomConst}\\
\dot{S}^{\mu\nu} & =0.\label{eq:TachyonSpinAngMomConst}
\end{align}
 The condition \eqref{eq:RefFourMomSpinTensorZero}, $p_{\refvalue,\mu}S_{\refvalue}^{\mu\nu}=0$,
now gives 
\begin{equation}
\mu c\,S_{\refvalue}^{z\nu}=0,\label{eq:Tachyon4MomSpinRelationship}
\end{equation}
 so the reference value of the system's spin tensor is 
\begin{equation}
S_{\refvalue}^{\mu\nu}=\begin{pmatrix}0 & \tilde{S}_{\refvalue,x} & \tilde{S}_{\refvalue,y} & 0\\
-\tilde{S}_{\refvalue,x} & 0 & S_{\refvalue,z} & 0\\
-\tilde{S}_{\refvalue,y} & -S_{\refvalue,z} & 0 & 0\\
0 & 0 & 0 & 0
\end{pmatrix}^{\mathclap{\mu\nu}}.\label{eq:TachyonRefSpinTensor}
\end{equation}
 The system's spin-squared scalar \eqref{eq:4DSpinSquaredInvariantFrom3Vecs}
and spin-squared pseudoscalar \eqref{eq:Def4DDualSpinSquaredInvariant}
have respective values 
\begin{align}
s^{2} & =S_{\refvalue,z}^{2}-\tilde{S}_{\refvalue,x}^{2}-\tilde{S}_{\refvalue,y}^{2},\label{eq:TachyonSpinSquared}\\
\tilde{s}^{2} & =0,\label{eq:TachyonDualSpinSquared}
\end{align}
 and the reference value of the system's Pauli-Lubanski pseudovector
\eqref{eq:PauliLubanski4VecFromSpin} is 
\begin{equation}
W_{\refvalue}^{\mu}=\mu c\parens{S_{\refvalue,z},\tilde{S}_{\refvalue,y},-\tilde{S}_{\refvalue,x},0}^{\mu}.\label{eq:TachyonRefPauliLubanski}
\end{equation}

We next consider the little group of orthochronous Lorentz transformations
that preserve the value of the reference four-momentum \eqref{eq:TachyonRef4Mom},
$p_{\refvalue}^{\mu}\defeq\parens{0,0,0,\mu c}^{\mu}$. As usual,
these little-group transformations parametrize the set of all states
that share that same four-momentum, so they include rotations around
the $z$ axis as well as Lorentz boosts along the $x$ and $y$ directions.

If the system is to have a compact set of states at any fixed four-momentum,
then its spin tensor \eqref{eq:TachyonRefSpinTensor} and Pauli-Lubanski
pseudovector \eqref{eq:TachyonRefPauliLubanski} must be invariant
under these noncompact Lorentz transformations. However, we see right
away that $W_{\refvalue}^{\mu}$ transforms nontrivially under Lorentz
transformations along the $x$ or $y$ directions if any of its components
are nonzero, so our system's phase space at fixed four-momentum can
be compact only if all the components of $W_{\refvalue}^{\mu}$ vanish:
\begin{equation}
\left.\begin{aligned}S_{\refvalue,z} & =0,\\
\tilde{S}_{\refvalue,x} & =0,\\
\tilde{S}_{\refvalue,y} & =0.
\end{aligned}
\quad\right\} \label{eq:TachyonRemainingSpinComponentsZero}
\end{equation}
 The tachyon's spin tensor and Pauli-Lubanski pseudovector therefore
vanish identically, 
\begin{equation}
\left.\begin{aligned}S^{\mu\nu} & =0,\\
W^{\mu} & =0,
\end{aligned}
\quad\right\} \label{eq:TachyonSpinTensorPauliLubanskiZero}
\end{equation}
 so 
\begin{equation}
\left.\begin{aligned}s^{2} & =0,\\
W^{2}\defeq w^{2} & =0,
\end{aligned}
\quad\right\} \label{eq:TachyonSpinSquaredPauliLubanskiSquaredZero}
\end{equation}
 and we see that a tachyon cannot have any intrinsic spin at all.

\subsection{The Vacuum}

Finally, we consider the case in which $p_{\refvalue}^{\mu}=0$, meaning
that the system's four-momentum vanishes for all the system's possible
states: 
\begin{equation}
p^{\mu}=0.\label{eq:VacuumVanishing4Mom}
\end{equation}
 The system then has no energy or momentum. The kinetic term $p_{\mu}\dot{X}^{\mu}$
in the system's action functional \eqref{eq:ActionFunctionalWithSpin}
vanishes, and we do not get a meaningful equation describing the behavior
of $X^{\mu}\parens{\lambda}$. The system's orbital angular momentum
vanishes, 
\begin{equation}
L^{\mu\nu}=0,\label{eq:VacuumVanishingOrbAngMom}
\end{equation}
 and its spin angular momentum is conserved: 
\begin{equation}
\dot{S}^{\mu\nu}=0.\label{eq:VacuumSpinTensorConst}
\end{equation}

The little group of orthochronous Lorentz transformations that leave
$p_{\refvalue}^{\mu}=0$ invariant consists of \emph{all} orthochronous
Lorentz transformations, and so the only way to obtain a compact phase
space at fixed four-momentum is for the spin tensor to vanish for
all the system's states: 
\begin{equation}
S^{\mu\nu}=0.\label{eq:VacuumVanishingSpin}
\end{equation}
 We conclude that our system is entirely devoid of energy, momentum,
and angular momentum, and therefore describes an empty vacuum.

\section{Conclusion}

In this paper, we reviewed a general method for making the standard
Lagrangian formulation manifestly covariant. We employed this framework
to develop a classical counterpart of Wigner's classification of quantum
particle-types in terms of the structure of the orthochronous Poincaré
group. We also showed that classical massless particles with spin
exhibit a novel manifestation of gauge invariance, and used the massless
limit to derive a classical version of the Higgs mechanism.

An interesting way to extend our approach is to consider phase spaces
that provide transitive group actions of the \emph{full} Poincaré
group, including time-reversal transformations \eqref{eq:TimeReversalTransf}.
This generalization does not affect our analysis of tachyons or of
the vacuum, which do not feature a definite sign for $p^{t}$. But
in the case of a system with non-negative mass, $m\geq0$, enlarging
the system's phase space so that it provides a transitive action of
the full Poincaré group means doubling the phase space to include
``negative-energy'' states with $p^{t}<0$. Because the four-momentum
$p^{\mu}$ is timelike or null when $m\geq0$, we know from \eqref{eq:CausalPropertySignTemporalComponentTimelikeNull4Vecs}
that the sign of $p^{t}$ is invariant under all \emph{physically
realizable} Lorentz transformations, which are smoothly connected
with the identity transformation and therefore do not include time-reversal
transformations. Hence, a system with $m\geq0$ cannot evolve from
states with $p^{t}>0$ to states with $p^{t}<0$ or vice versa. We
are therefore free to define the physical energy of the additional
$p^{t}<0$ states to be $E\defeq-p^{t}c>0$, and regard them as states
not of our original particle, but of its corresponding antiparticle.
In this way, we can classically unify particles with their antiparticles.
\begin{acknowledgments}
J.\,A.\,B. has benefited tremendously from personal communications
with Howard Georgi, Andrew Strominger, David Griffiths, David Kagan,
David Morin, Logan McCarty, Monica Pate, and Alex Lupsasca.
\end{acknowledgments}

\bibliographystyle{2_home_jacob_Documents_Work_My_Papers_Manifestl___ant_Lagrangians__2020__custom-abbrvunsrturl}
\bibliography{0_home_jacob_Documents_Work_My_Papers_Bibliography_Global-Bibliography}

\end{document}